\documentclass[sigconf]{acmart}
\AtBeginDocument{%
  }

\copyrightyear{2026}
\acmYear{2026}
\setcopyright{cc}
\setcctype{by}
\acmConference[CHI '26]{Proceedings of the 2026 CHI Conference on Human Factors in Computing Systems}{April 13--17, 2026}{Barcelona, Spain}
\acmBooktitle{Proceedings of the 2026 CHI Conference on Human Factors in Computing Systems (CHI '26), April 13--17, 2026, Barcelona, Spain}
\acmPrice{}
\acmDOI{10.1145/3772318.3790694}
\acmISBN{979-8-4007-2278-3/2026/04}

\usepackage{enumitem}
\usepackage{xcolor}
\usepackage{multirow}
\usepackage{graphicx}
\usepackage{stfloats}

\usepackage{fancyvrb}
\usepackage{fvextra}

\newcommand{\sysname}{Iconix}

\begin{document}
\captionsetup[figure]{name={Figure}}

\title{Iconix: Controlling Semantics and Style in Progressive Icon Grids Generation}

\author{Zhida Sun}
\orcid{0000-0003-4689-986X}
\affiliation{
  \institution{CSSE, Shenzhen University}
  \city{Shenzhen}
  \country{China}
}
\email{zhida.sun@szu.edu.cn}

\author{Xiaodong Wang}
\orcid{0009-0004-3942-3980}
\affiliation{%
  \institution{CSSE, Shenzhen University}
  \city{Shenzhen}
  \country{China}
}
\email{zxx70611@gmail.com}

\author{Zhenyao Zhang}
\orcid{0009-0004-9579-1454}
\affiliation{%
 \institution{CSSE, Shenzhen University}
 \city{Shenzhen}
 \country{China}}
\email{zhenyaoz66@gmail.com}

\author{Min Lu}
\orcid{0000-0002-8464-0990}
\affiliation{
 \institution{Shenzhen University}
 \city{Shenzhen}
 \country{China}}
 \email{lumin.vis@gmail.com}

\author{Dani Lischinski}
\orcid{0000-0002-6191-0361}
\affiliation{
  \institution{Hebrew University of Jerusalem}
  \city{Jerusalem}
  \country{Israel}}
\email{danix@cs.huji.ac.il}

\author{Daniel Cohen-Or}
\orcid{0000-0001-6777-7445}
\affiliation{
    \institution{Tel Aviv University}
  \city{Tel Aviv}
  \country{Israel}
  }
\email{cohenor@gmail.com}

\author{Hui Huang}
\authornote{Corresponding author}
\orcid{0000-0003-3212-0544}
\affiliation{
  \institution{CSSE, Shenzhen University}
  \city{Shenzhen}
  \country{China}
  }
\email{hhzhiyan@gmail.com}

\renewcommand{\shortauthors}{Z. Sun, X. Wang, Z. Zhang, M. Lu, D. Lischinski, D. Cohen-Or, and H. Huang}

\begin{abstract}
    Visual communication often needs stylistically consistent icons that span concrete and abstract meanings, for use in diverse contexts. We present \sysname, a human-AI co-creative system that organizes icon generation along two axes: \emph{semantic richness} (what is depicted) and \emph{visual complexity} (how much detail). Given a user-specified concept, \sysname\ constructs a semantic scaffold of related analytical perspectives and employs chained, image-conditioned generation to produce a coherent style of exemplars. Each exemplar is then automatically distilled into a progressive sequence, from detailed and elaborate to abstract and simple. The resulting two-dimensional grid exposes a navigable space, helping designers reason jointly about figurative content and visual abstraction. A within-subjects study ($N=32$) found that compared to a baseline workflow, participants produced icon grids more creatively, reported lower workload, and explored a coherent range of design variations. We discuss implications for human-machine co-creative approaches that couple semantic scaffolding with progressive simplification to support visual abstraction. 
\end{abstract}



\begin{teaserfigure}
  \centering
  \includegraphics[width=\textwidth]{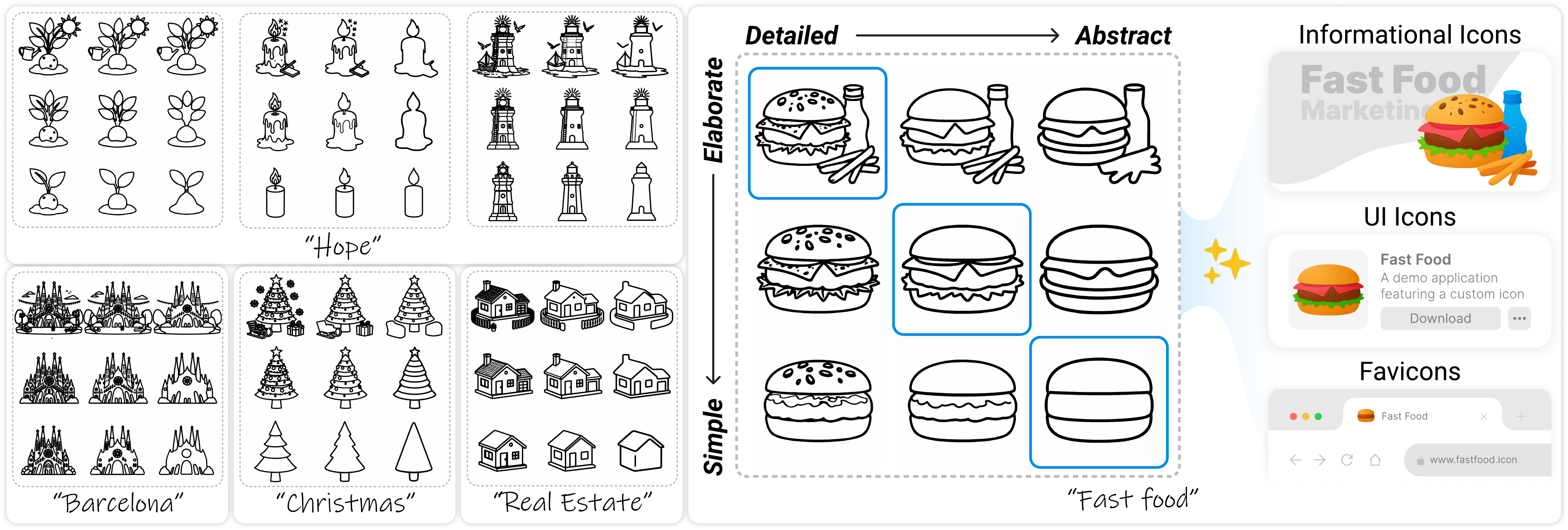}
  \caption{\sysname\ generates icon grids from a single input concept.
  The system automates the transition from abstract concepts to a coherent icon continuum by decomposing semantic relationships and performing progressive visual simplification. The result is a structured grid of icons tailored to diverse application scenarios.
  }
  \Description{The figure presents multiple icon grids generated from different concepts, including ``Hope'', ``Barcelona'', ``Christmas'', ``Real Estate'', and ``Fast food''. Each grid contains hand-drawn icons arranged from elaborate to simple forms, and from detailed to abstract visual styles. On the right, several polished output examples show how the generated icons can be used as informational graphics, UI icons, and website favicons.}
  \label{fig:teaser}
\end{teaserfigure}

\maketitle

\section{Introduction}

Icons are a ubiquitous and fundamental element of modern user interfaces, serving as compact, language-independent visual representations~\cite{10.1145/3664647.3681057, 10.1007/978-3-319-20886-2_3}, supporting interaction, and conveying a product’s visual identity~\cite{hicks2011icon, 10.5555/2692726, 10.1145/2556288.2557408, 10.1145/3313831.3376618}.
Beyond mere decorations, effective icons must distill complex ideas into recognizable forms~\cite{COLLAUD2022102290},
while adhering to constraints such as stylistic consistency, brand identity, or technical rendering specifications~\cite{10.1145/3404983.3405518}.
Contemporary design practice demands that icons remain effective across diverse contexts, from ensuring legibility across a spectrum of devices (e.g.,~\cite{https://doi.org/10.1002/eng2.12577, 10.1145/3697355.3697360}) to providing accessibility for users with varied capabilities (e.g.,~\cite{10.1007/978-3-030-78108-8_10}). 
Consequently, the designer's task has expanded from creating a single icon to engineering a set of representations for a concept, each maintaining a stable semantic core while offering stylistic and structural variations (see Figure~\ref{fig:teaser}).

Recent advances in generative AI, particularly large-scale text-to-image (T2I) models, offer compelling opportunities to accelerate the icon design workflow (e.g., \cite{10.1145/3531065, 10.1145/3664647.3681057}). Yet a significant gap remains between the expressive power of these models and the nuanced needs of icon design.
The core challenges are twofold. 
First, achieving precise semantic control is difficult, as models often struggle to generate the specific imagery that captures a designer's intent, particularly for abstract concepts~\cite{Liao_Chen_Fu_Du_He_Wang_Han_Zhang_2024, 10.1145/3706598.3713683}. 
Second, semantic and stylistic controls are tightly entangled in these models. Attempts to adjust content often disrupt stylistic coherence, while enforcing strict stylistic constraints can suppress necessary semantic variation~\cite{10.1145/3706598.3714292}.
This tension between controlling what an icon depicts (semantics) and its appearance (style) traps designers in a tedious, trial-and-error loop, limiting the potential of generative tools for streamlining the creation of cohesive icon series.

To bridge this gap, our research introduces a computational method that enables designers to generate style-consistent and semantically organized icons from a single conceptual input.
While not intended to replace professional vector editors or end-to-end design pipelines, our goal is to help designers move beyond the tedious task of creating individual icons and instead focus on exploring a wide range of creative possibilities, especially to facilitate rapid ideation and the exploration of abstraction.
To facilitate this exploration, we introduce the term \textit{icon grid} to define a structured continuum of style-consistent representations derived from a single semantic concept, organized along two axes of semantic richness and visual complexity.
This grid represents the full spectrum of an icon's potential form, ranging from concrete, illustrative depictions to abstract, symbolic forms~\cite{10.1145/1362550.1362591, 10.1007/978-3-319-91238-7_11}.
To this end, our approach mirrors a progressive refinement procedure, employing an additive process to systematically reassemble semantic components and ensure a coherent progression of detail across the final icon grid.

We operationalize this method in \sysname, a human-AI co-creative system designed to streamline the icon grid design workflow.
The process begins with a user's initial theme. 
To broaden creative exploration, \sysname\ first leverages large language models and knowledge bases to help the designer refine the input into a core imagery concept. 
This concept is then expanded into a ``semantic scaffold'' built from structured semantic relations, allowing the designer to consider different facets of the concept before generation. 
Following this scaffold, the system employs a chained, image-conditioned generation process to produce a set of stylistically consistent exemplars. 
Each exemplar is then automatically simplified into a structured continuum of icons that progress from an elaborated, detailed version to a simple, abstract representation for the designer's review.
We explored the implications of this approach through a within-subjects study with 32 participants. Our findings provide indicative evidence that \sysname\ supports structured semantic–visual exploration and reduces the manual effort inherent in creating coherent icon grids.
We conclude by distilling participant feedback into qualitative insights that inform the design of visual abstraction workflows and human–AI co-creative systems.
Our primary contributions are threefold:
\begin{itemize}[leftmargin=*, noitemsep]
    \item We formalize a computational method for generating progressive icon grids, coupling generative AI models to systematically expand a single concept into a structured continuum of representations along the two-dimensional axis of semantic richness and visual complexity.
    \item We present the design and implementation of \sysname, a human-AI co-creative system that operationalizes our method. It automates the laborious aspects of icon grid creation while preserving designer-in-the-loop creative control.
    \item We provide empirical validation to demonstrate that our approach streamlines the design of coherent icon grids and fosters creative exploration. Based on these findings, we offer insights on designing visual abstraction and human-machine co-creation.
\end{itemize}

\section{Related Work}

\subsection{Controlling Icon Semantics and Style}

Icon design is a fundamental aspect of Human-Computer Interaction, tasked with communicating complex semantics through compact visual forms to ensure usability and recognizability~\cite{BUHLER2022102816, COLLAUD2022102290}.
A fundamental tension in icon design lies in balancing semantic clarity with stylistic consistency~\cite{Hou24022025, 10.1007/978-3-030-78108-8_10, https://doi.org/10.1002/hfm.70006}.
Semantics refers to the meaning or content conveyed by an icon, such as representing a ``search'' function or a ``calendar'' event, while style encompasses visual attributes like color, shape, and line art that ensure consistency and aesthetic appeal~\cite{COLLAUD2022102290}.
Generally, an icon should clearly communicate its intended meaning (semantics) while adhering to a specific visual language (style) that ensures it feels part of a cohesive whole.
This challenge has been a long-standing focus of computational design tools.

To address this challenge, early computational systems provided designers with explicit controls, often focusing on either semantic composition or stylistic application.
Systems like ICONATE~\cite{10.1145/3313831.3376618}, for instance, focused on semantic control by allowing designers to compose new, complex icons from existing graphical primitives based on a textual query. 
This approach offered a high degree of control over the icon's content but was limited by the creative space of the predefined elements. Conversely, other systems have focused exclusively on style. GAN-based colorization tools like FlexIcon~\cite{10.1145/3581783.3612182} and the work of Sun et al.~\cite{10.1145/3343031.3351041} offer fine-grained control over an icon's color scheme but treat its semantic structure as a fixed input provided by the user. 
These tools excel at applying a consistent style but do not assist in generating or varying the underlying semantic content.

More recent work has approached this problem through the lens of disentangled representation learning, which aims to create models where latent factors independently control specific visual attributes. 
In icon design, this was demonstrated by IconGAN~\cite{10.1145/3503161.3548109}, which used orthogonal ``app'' (semantic) and ``theme'' (style) labels with a dual-discriminator architecture to achieve disentangled control over the generated output. 
While these methods advanced the state of controllable generation, they were often limited by their reliance on discrete labels or a focus on a single dimension of control, rather than supporting a fluid exploration of the trade-offs between semantics and style.
Our work addresses this gap by moving beyond the paradigm of single-output generation to the creation of a structured spectrum of design alternatives. This provides direct computational support for a core, yet previously unaddressed, aspect of the icon design workflow, that is, the need to explore and select from a range of stylistically consistent and semantically related options for a given concept.

\subsection{Progressive Visual Content Abstraction}

An icon functions as a representation that connects a visual form to its underlying meaning, a connection that exists on a spectrum from concrete depiction to abstract symbol.
The effectiveness of an icon is heavily influenced by its level of abstraction or concreteness, a topic with a long history of study in HCI~\cite{mcdougall2000exploring,10.1145/3192975.3192980, Goetz17112024}. 
Research has consistently shown that the trade-off between these two properties impacts user performance~\cite{mcdougall2000exploring,10.1145/3192975.3192980, Goetz17112024}, which means that designers must often create multiple representations of the same concept for different contexts, a manual process analogous to creating multiple Levels of Detail (LoD) in computer graphics, which is inefficient and ill-suited for dynamic contexts.

To address this, computational methods have explored ways to control the visual attributes of icons. 
One direction is language-driven geometric editing, which establishes direct mappings between natural language and geometric attributes for fine-grained modifications. 
For example, Xu et al.~\cite{xu2024creatinglanguagedrivenspatialvariations} introduced a technique that parses instructions like ``shorten the bus length by 30\%'' into geometric constraints via a domain-specific language. While this allows for precise, interpretable edits, it relies on predefined geometric primitives and is primarily focused on modifying an existing icon's structure rather than generating a range of new ones from a single concept.
More recent direction involves the continuous modulation of visual attributes via a generative model’s latent space. 
By interpolating or traversing latent vectors, specific attributes of an image can be adjusted smoothly. Research in this area has demonstrated three main types of control.
First, for \textit{structural and detail control}, Wang et al.~\cite{11094816} proposed a hierarchical vectorization framework that decomposes images into geometric and textural layers, while CLIPascene~\cite{10378256} utilizes multi-scale decomposition and CLIP alignment to enable continuous adjustment from simple line art to photorealistic renderings. 
Second, for \textit{semantic attribute and style modulation}, Li et al.~\cite{10.1145/3696410.3714912}, for instance, introduced an intensity coefficient in a diffusion model to smoothly adjust abstract features like ``a sense of safety'', while EmotiCrafter~\cite{Dang_2025_ICCV} maps emotional descriptions to a valence-arousal space for targeted visual generation. 
Third, to make these powerful controls accessible, HCI research has explored \textit{interactive modulation}. AdaptiveSliders~\cite{10.1145/3706598.3714292}, for example, exposes these latent dimensions to the user through sliders, allowing them to intuitively adjust attribute intensities with real-time feedback, thereby enhancing the efficiency of human-AI collaboration.

The predominant focus of these methods is the modification or fine-tuning of a single, existing visual instance. 
They lack a systematic way to deconstruct a visual concept into its constituent semantic primitives and then progressively reassemble them to generate a structured icon continuum from a unified source. 
Our work addresses this gap by introducing a method that computationally generates a structured spectrum of design alternatives from detailed, concrete depictions to simple, abstract forms. 
Instead of editing a single artifact, this approach provides direct support for the creation of an entire adaptive hierarchy for a concept, a core yet previously unaddressed challenge in the design workflow.

\subsection{Creating Icons with Generative AI}

The recent proliferation of large-scale generative AI has introduced a new paradigm for icon creation, enabling the synthesis of novel icons from scratch and offering powerful new avenues for creative support.
Early generative approaches relied on Generative Adversarial Networks to produce \textit{pixel-based icons}, as exemplified by IconGAN~\cite{10.1145/3503161.3548109}. 
More recently, diffusion models have become the state-of-the-art, offering higher fidelity and better prompt adherence. IconDM~\cite{10.1145/3664647.3681057}, for instance, uses a diffusion model to generate new icons for different semantic concepts that conform to a user-provided style reference. 
In parallel, other research has focused on generating \textit{vector-based icons}, which are friendly for design workflows due to their scalability. 
IconShop~\cite{10.1145/3618364} employs an autoregressive transformer to produce vector icons directly from textual descriptions, demonstrating a different technical pathway for generative icon creation.

Beyond pure generation, research has explored how to integrate these capabilities into design workflows through human-AI collaboration.
Some tools, like Auto-Icon$+$~\cite{10.1145/3531065}, focus on downstream integration, automatically converting icon images into fonts and generating descriptive labels to enhance code accessibility. 
Other systems position the AI as an evaluative partner. 
EvIcon~\cite{https://doi.org/10.1111/cgf.14924}, for example, uses a Vision-Language Model to provide designers with feedback on the usability and semantic clarity of the icons they are manually creating.
These systems excel at accelerating the initial ideation phase by producing a wide array of visual concepts.

Despite these advances, prior methods have largely focused on either initial ideation or final implementation, leaving a critical gap in supporting the intermediate, yet highly laborious, stage of scaling and versioning a design concept for adaptive use.
Our work addresses this gap by introducing a method that uses pixel-based generation to create a systematically organized grid of icon representations, automating the complex structural task of generating an entire adaptive hierarchy. This approach provides direct support for a core, yet previously unaddressed, aspect of the human-AI collaborative workflow for icon design.

\section{Formative Study}

To ground our system's design in core design strategies and make them accessible to general users, we conducted a formative study with professional designers.
Our goal was to distill expert workflows and understand where even professionals struggle with effort and consistency, enabling us to operationalize the workflows and target computational support to lower the barrier for general users.

\subsection{Method}

Aligning with prior HCI research (e.g.,~\cite{10.1145/3313831.3376618}) that relies on focused expert samples to ground system design in domain-specific practices, we conducted semi-structured interviews with three professional designers (2 female, 1 male; ages 32-43), each with over 10 years of experience in graphic design, UI/UX, and illustration. 
Participants were recruited from a local design community and purposefully sampled to represent a diverse range of icon creation contexts, spanning from web application interfaces to scientific publications. While small, this sample size was appropriate for our exploratory goal, as we sought to surface qualitative patterns in expert workflows rather than obtain statistical generalizations. 
Each interview lasted 45–60 minutes and was audio-recorded, transcribed verbatim, and anonymized (see Appendix B for the full question list).
The interviews focused on three main areas:
\begin{itemize}[leftmargin=*, noitemsep]
    \item Current workflows: We asked designers to walk through their typical process for sourcing or creating icons, from initial concept to final asset, including the tools they use.
    \item Core challenges: We probed the specific difficulties they face, particularly concerning the trade-off between semantic accuracy and visual characteristics, the creation of icons for abstract concepts, and the need to adapt icons for different devices and contexts.
    \item Unmet needs and tooling gaps: We explored their ideal features in a design tool, asking them to envision a function that could solve their most significant pain points. We also presented the core concept of our progressive generation model to gauge its potential value and gather initial feedback.
\end{itemize}

\subsection{Findings}\label{sec:findings}

We analyzed the interview data using deductive thematic analysis based on Braun and Clarke’s framework~\cite{Braun01012006}. 
Two researchers independently coded the transcripts, specifically identifying statements related to abstraction, figurative representation, and semantic–visual control. 
These codes were consolidated through iterative discussion and synthesized into higher-level themes. 
We observed a strong convergence of perspectives, with core workflow challenges recurring consistently across all three expert accounts.
This consistency provided a foundation for identifying the primary design opportunities reported below.
The coding scheme and codebook are provided in the supplementary materials.

\paragraph{Achieving Abstraction through Progressive Simplification} 
A consistent pattern emerged in how designers manage visual complexity. 
When designing minimalist icons, professional designers typically adopt a subtractive workflow as one option. This process entails progressive simplification, where the designer initiates the work with a detailed, concrete representation and systematically removes visual elements or simplifies shapes until only the essential form remains.

One expert (E1, Female, 43) with 20 years of experience described this as a natural process of moving ``\textit{from real to simplified}'' and noted that a tool providing ``\textit{ready-made progressive choices}'' would be highly intuitive and time-saving. 
This strategy is especially critical when adapting icons for various sizes, as maintaining legibility on smaller scales requires the intentional removal of details. 

\paragraph{Finding the Right Figurative Representation as a Core Bottleneck}
A major source of frustration for designers is the frequent mismatch between their intended meaning and the icons available to them, a challenge the interviewed professionals referred to as a ``\textit{semantic gap}''. 
These experts repeatedly described having to settle for a ``\textit{relatively suitable}'' choice rather than the ideal one, which forces them into an inefficient workflow of searching multiple libraries, attempting to combine existing assets, or designing from scratch as a last resort.
This problem is amplified when designing for abstract concepts. 
For these ideas, a shared mental image is often weak or non-existent, making any visual representation ambiguous and open to interpretation. 
The professionals reported that for these concepts, existing options were often ``\textit{too complex}'' or ``\textit{too abstract}'', failing to align with their mental model. 
The core difficulty lies in translating an intangible idea into a clear visual metaphor. 
This task relies heavily on the individual experience and skill of the designer, often lacks a clear starting point, and can lead to creative block, highlighting a lack of systematic support in current tools for exploring ``\textit{what to draw}''.

\paragraph{The Need for Decoupled Semantic and Stylistic Control}
While designers seek inspiration, they require a high degree of control to ensure the final output is consistent with design guidelines and project requirements. 
A recurring desire was for tools that could generate a stylistically unified icon continuum.
These experts explicitly requested independent control over an icon's conceptual and visual properties. As one expert (E2, Male, 40) articulated, an ideal tool would allow them to adjust both the amount of conceptual detail (e.g., choosing between a simple ``\textit{tree}'' versus a ``\textit{decorated Christmas tree}'') and stylistic detail (e.g., controlling line weight and form). 
This need is driven by the dual challenge of context and communication, as adapting for media size requires tuning visual complexity to ensure legibility, while tailoring content for specific audiences demands the independent adjustment of semantic richness.
Currently, this process of determining ``\textit{how to represent it}'' for different contexts is a point of friction in the manual workflow, requiring significant effort to ensure consistency across multiple representations.

\subsection{System Design Requirements}

Based on these findings, we formulated a set of core design requirements (DR) to guide the development of a system that addresses the identified issues and aligns with the practices of professional designers.

\begin{enumerate}[leftmargin=*, noitemsep, label={\textbf{DR\arabic*}}]
    \item Support Semantic Exploration. 
    The system needs to assist designers in the initial ideation phase for given inputs. It should help explore ``\textit{what to draw}'' by generating a diverse range of visual ideas that are clearly connected to the intended meaning. This support is particularly valuable for overcoming the creative block associated with abstract concepts, where visual metaphors are not obvious.
    \item Automate Progressive Simplification. 
    The system needs to mirror the designers' subtractive workflow of starting with a more detailed concept and simplifying it. It should automatically generate a structured continuum of representations for a concept, ranging from detailed to abstract, to support the need for icons that are adaptable to different contexts.
    \item Provide Independent Control over Semantics and Style. 
    The system needs to provide designers with independent control over an icon's semantic complexity (what is depicted) and its visual complexity (the level of detail). This addresses the need to explore ``\textit{how to represent it}'' for different scenarios and moves beyond a scattered collection of options to a structured exploration of design trade-offs.
    \item Ensure Stylistic Cohesion. 
    The system needs to generate all representations within an icon continuum with a consistent visual style to ensure a unified visual language for communication. This addresses the critical need for a stylistically unified set that aligns with design requirements, which is currently a labor-intensive manual task.
\end{enumerate}

\section{\sysname}\label{sec:system}

\subsection{Design Consideration}

Informed by the findings and system design requirements from our formative study, we established a set of design considerations to guide the system's functionality and user interface. 
These considerations explain the reasoning behind our specific design choices in translating our goals into a functional, human-centered system.

\begin{enumerate}[leftmargin=*, noitemsep, label={\textbf{C\arabic*}}]
    \item Scaffolding Semantic Exploration (DR1): 
    To support designers in exploring ``\textit{what to draw}'', we consider how to present a wide range of semantic possibilities without overwhelming them. 
    Our design choice is to present a list of scored concepts for quick comparison and selection, and a tree-structure visualization of semantic relations for deeper, more exploratory ideation.
    \item Making Abstraction Transparent and Controllable (DR2, DR3): To support the designer's task of determining ``\textit{how to represent it}'' for different contexts, we consider how to make the automated abstraction process both transparent and controllable. 
    Our design solution incorporates two parts. First, a progressive simplification sequence makes the process transparent by showing the full spectrum of options for user supervision. 
    Second, a grid organized by two orthogonal axes, which captures the two-dimensional nature of the problem, allowing for direct navigation and overall control of the entire creation space.
    \item Ensuring Stylistic Cohesion for Generated Visual Content (DR4):
    A key challenge in generative workflows is maintaining stylistic consistency. 
    To address the trade-off between generative diversity and cohesion, our design choice is to root all variations in a series of chained, pixel-based exemplars. These act as a stylistic anchor, ensuring that all subsequent, simpler representations derived from them share a consistent visual identity, thus fulfilling the goal of stylistic cohesion from the outset.
\end{enumerate}

\subsection{System Design}

\begin{figure*}[t]
  \centering
  \includegraphics[width=\textwidth]{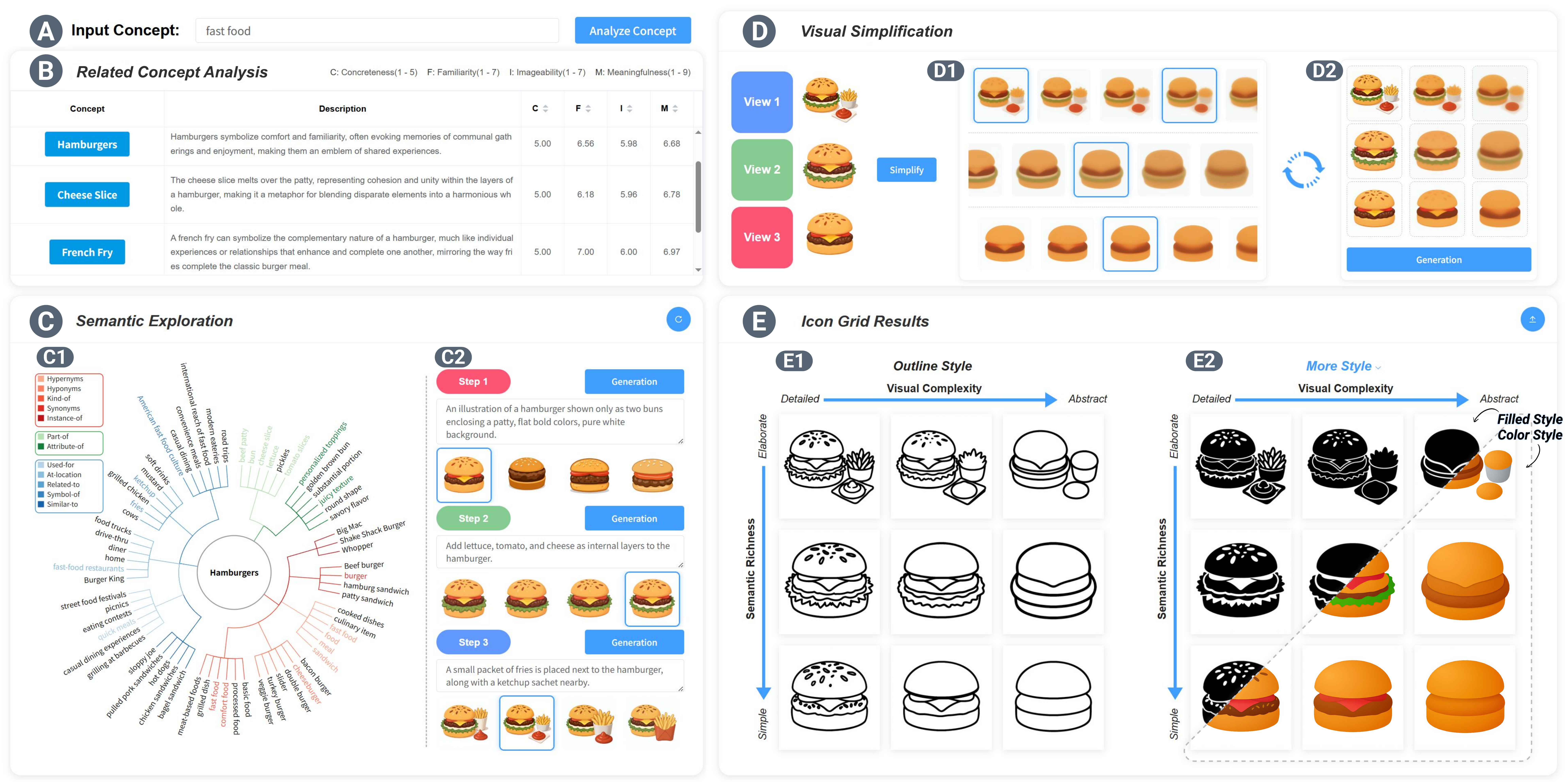}
  \caption{The \sysname's interface showcases an example of generated results for ``fast food''. The system comprises five coordinated modules: (A) Input Concept; (B) Related Concept Analysis (showing candidate ratings); (C) Semantic Exploration (visualizing associations and prompts); (D) Visual Simplification; and (E) the Dual-Axis Icon Grid for comparing design trade-offs.
  }
  \label{fig:interface}
  \Description{The figure displays the Iconix interface arranged into five labeled modules for the example concept ``hamburger''. Module A shows a text input box with the entered concept. Module B contains a table of related concepts with descriptive explanations and numeric attribute ratings. Module C includes a semantic exploration view with a radial relationship map on the left and a step-by-step prompt panel with corresponding hamburger illustrations on the right. Module D presents visual simplification options, showing multiple image variations across three views and a panel for generating simplified versions. Module E shows two icon grids organized along axes of visual complexity and semantic richness, comparing outline and filled styles across detailed to abstract representations.}
\end{figure*}

\subsubsection{A Dual-Axis Grid for Icon Representation}

In the realm of user interface design, the efficacy of an icon often hinges on its visual representation. 
While icons inherently serve as a simplified form of visual identity, the degree and nature of this simplification can vary significantly. 
To address this, we propose a two-dimensional matrix for systematically organizing icon representations, offering a framework for both analysis and design exploration.

The primary interface for exploring the generated icon continuum is a dual-axis grid, which organizes the icons along two orthogonal dimensions: semantic richness and visual complexity.

\paragraph{Semantic Richness \texttt{\footnotesize (vertical axis)}} This axis controls what is depicted in the icon by managing the number of meanings, associations, and semantic features connected to a concept. It represents a continuum of conceptual detail, ranging from a single, core element to a rich composition of multiple related elements.
\begin{itemize}[leftmargin=*, noitemsep]
    \item High Semantic Richness: The icon is a composition of multiple semantic features, evoking a wider range of connections. For example, an icon depicting an ``apple'' with a leaf, a flower, and a drop of dew enhances the semantic features, suggesting associations like nature, growth, freshness, and health.
    \item Low Semantic Richness: The icon represents a single, decontextualized concept with a limited set of immediate associations. For example, an icon showing just an apple represents the core concept of ``apple'' (fruit, food).
\end{itemize}
    
\paragraph{Visual Complexity \texttt{\footnotesize (horizontal axis)}} This axis controls how the selected semantic content is rendered. It represents a spectrum of visual abstraction, from detailed, naturalistic forms to simplified, geometric shapes.
\begin{itemize}[leftmargin=*, noitemsep]
    \item High Visual Complexity: At this end of the spectrum, the representation is highly illustrative. Icons here retain more detail, with curvilinear forms and irregular outlines reminiscent of their real-world counterparts.
    \item Low Visual Complexity: At this end, the representation is highly symbolic. Icons here have low visual complexity and are rendered using fundamental geometric primitives. This emphasizes minimalism and a more symbolic or modern aesthetic, optimized for legibility at small sizes.
\end{itemize}

\subsubsection{User Interface and Workflow}

The \sysname\ system integrates a multi-stage pipeline to support designers in creating progressive and style-consistent icon grids. 
The user interface (see Figure~\ref{fig:interface}) is structured into four main modules that guide the designer from initial concept exploration to the final selection of an icon grid.

\paragraph{Concept Analysis and Selection} 
The workflow begins when a designer inputs an initial concept. To support semantic exploration (C1), the system first analyzes the input and presents a structured candidate table of related concepts~\cite{10.1145/3313831.3376618, 10.1145/3313831.3376746}. This table includes descriptive information and ratings across four dimensions, i.e., Concreteness, Familiarity, Imageability, and Meaningfulness~\cite{martinez2024using, brysbaert2024moving}, to help users select the most appropriate design directions for further exploration.

\paragraph{Semantic Exploration}
To further expand creative potential, \sysname\ provides a semantic exploration module~\cite{10.1145/3706598.3713935,10.1145/3706598.3713683}. A tree-structure visualization displays the semantic relations extracted from knowledge bases, enabling designers to discover associative connections that extend beyond the literal meaning of their initial input~\cite{wang2025evaluatingnodetreeinterfacesai} (C1). The system then guides the user through a stepwise semantic progression process across three analytical perspectives
(i.e., Taxonomic, Constitutive, and Associative). This structured exploration enables them to gradually generate content that aligns with both their intended ideas and desired visual style (C3).

\paragraph{Visual Simplification}
In the visual simplification module, the system generates icons for the selected concept at different abstraction levels. To make the process transparent (C2), the icons are organized in a clear sequence based on progressive simplification~\cite{10.1145/3528223.3530068, 11094816}. This allows users to directly compare how varying levels of detail affect an icon's recognizability and adaptability~\cite{10.1145/3746059.3747631}. The interface uses color-coded layouts and similarity-based grouping to enhance clarity and facilitate this comparison. 

\paragraph{Dual-Axis Exploration and Style Refinement}
The system presents the complete generated icon candidates on a structured 2D canvas~\cite{mccloud1993understanding, 10378256}. The results are organized along two orthogonal axes (C2): visual complexity (ranging from detailed to abstract) and semantic richness (from elaborate to simple). This dual-axis representation allows designers to efficiently evaluate and select icon candidates that balance stylistic consistency with semantic clarity~\cite{10.1145/2556288.2557408}.
To support final design decisions (C3), this module ensures stylistic consistency across different visual variants, including outline, filled, and colorized versions~\cite{10.1145/3664647.3681057}. This enables designers to examine the style alignment across these major formats, informing further development.

\subsection{Implementation}

\subsubsection{Semantic Expansion for Concept Ideation}

\begin{figure*}[t]
  \centering
  \includegraphics[width=\textwidth]{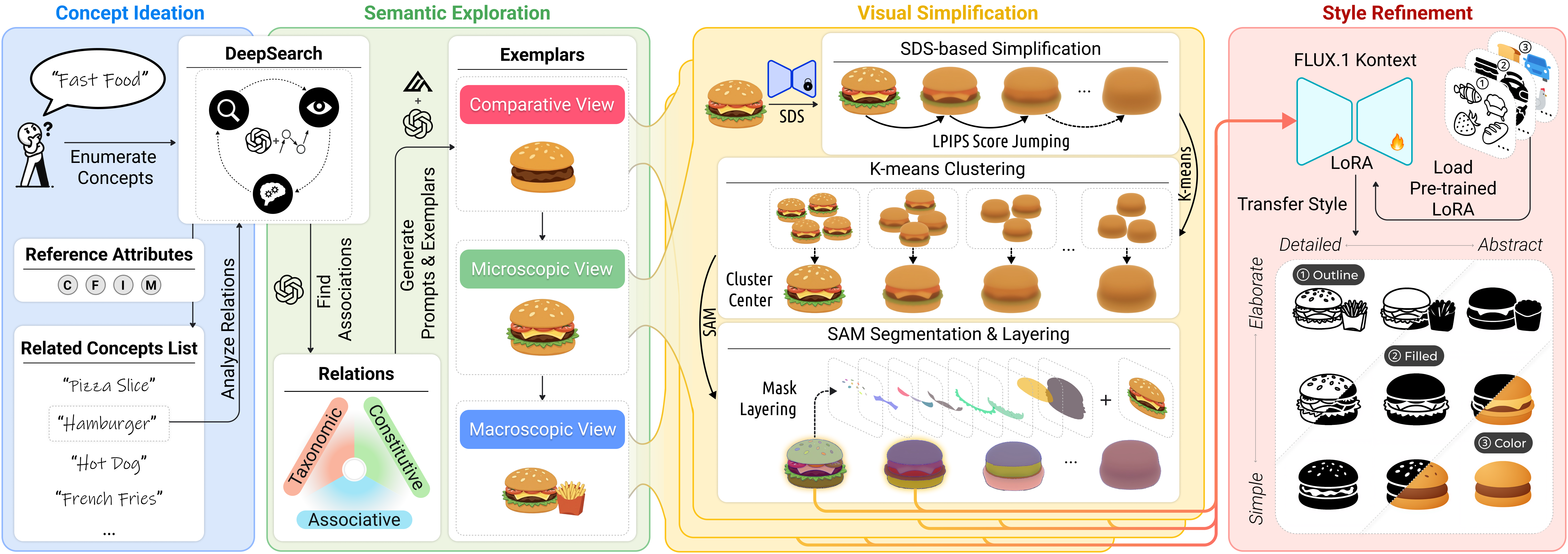}
  \caption{Overview of the \sysname\ pipeline. The pipeline consists of four stages: (a) Concept Ideation identifies related concepts from the input; (b) Semantic Exploration analyzes relations and constructs three-level prompts to generate image exemplars; (c) Visual Simplification transforms the exemplars into icon mask drafts of varying complexity; and (d) Style Refinement optimizes the drafts to generate multi-style icon grids.  }
  \label{fig:pipeline}
  \Description{The figure illustrates the Iconix pipeline organized into four color-coded sections. On the left, the Concept Ideation area shows a user providing the concept “fast food,” producing reference attributes and a list of related concepts. The Semantic Exploration section displays LLM-driven relation analysis together with exemplar images shown in comparative, microscopic, and macroscopic views. The Visual Simplification section shows multiple operations—SDS-based simplification, K-means clustering of hamburger images, and SAM-based segmentation with layered mask outputs—each producing increasingly abstract visual forms. On the right, the Style Refinement section depicts a model applying pre-trained LoRA styles to generate icon grids in outline, filled, and color styles arranged along axes of detail and abstraction.}
\end{figure*}

To address the challenge of visualizing abstract concepts, the first module of \sysname\ implements a computational pipeline for semantic expansion, shown in Figure~\ref{fig:pipeline}.
Instead of representing abstract concepts through compound visual designs~\cite{10.1145/3313831.3376618}, our process translates a single abstract input (e.g., ``hope'') into a ranked list of concrete, visualizable candidate concepts. The pipeline employs a multi-round iterative reasoning framework composed of three stages: searching, evaluation, and reasoning.

In the \raisebox{-.2ex}{\includegraphics[height=2ex]{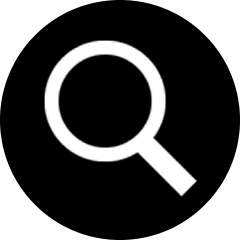}} searching stage, building on the semantic expansion approach from prior work~\cite{10.1145/3706598.3713683},
the system queries both ConceptNet 5~\cite{10.5555/3298023.3298212} and OpenAI's GPT-4o model\footnote{GPT-4o [\texttt{gpt-4o-2024-08-06}]: https://platform.openai.com/docs/models/gpt-4o} to acquire an initial set of semantically related objects. 
For the concept ``hope'', this stage might identify candidates like ``phoenix'', ``sunrise'', ``lighthouse'', or ``seed''. 
Each iteration refines the search by incorporating previously found concepts, progressively broadening the candidate pool.

In the \raisebox{-.2ex}{\includegraphics[height=2ex]{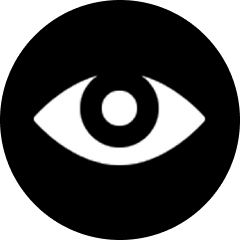}} evaluation stage, each candidate is assessed for its figurative potential. 
First, we utilize structured prompting~\cite{10.1145/3706598.3713683} to generate a qualitative interpretation that explains the semantic connection between the concrete object and the abstract concept. 
For example, the system rationalizes the association between ``seed'' and ``hope'' by explaining ``a seed embodies the promise of growth and new beginnings, symbolizing the hope inherent in potential''.
Second, we quantitatively score each candidate along four psycholinguistic attributes to modulate icon comprehensibility~\cite{martinez2024using, brysbaert2024moving}: \raisebox{-.2ex}{\includegraphics[height=2ex]{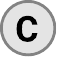}} Concreteness (tangibility), \raisebox{-.2ex}{\includegraphics[height=2ex]{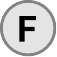}} Familiarity (frequency of encounter), \raisebox{-.2ex}{\includegraphics[height=2ex]{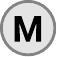}} Meaningfulness (interpretability), and \raisebox{-.2ex}{\includegraphics[height=2ex]{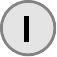}} Imageability (ease of visualization). 

Finally, in the \raisebox{-.2ex}{\includegraphics[height=2ex]{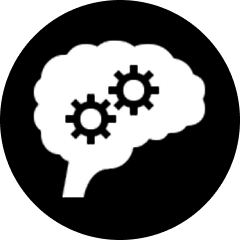}} reasoning stage, the system synthesizes the extracted semantic insights to filter and rank the candidate referents. 
Executing a multi-criteria ranking protocol, the reasoning engine prioritizes candidates via a three-step logic:
\begin{itemize}[leftmargin=*, noitemsep]
    \item Constraint Filtering removes abstract nouns (e.g., ``optimism''), superordinate categories (e.g., ``plant''), and intangible actions; 
    \item Attribute Thresholding retains candidates with high visual recognizability (Concreteness $\ge$ 4/5, Familiarity \& Imageability $\ge$ 5/7) and strong metaphorical connection (Meaningfulness $\ge$ 6/9);
    \item Weighted Ranking sorts survivors by aggregated attribute scores to identify the top ones.
\end{itemize}
This iterative process incorporates new concepts recursively, terminating when the top-5 candidate set stabilizes or reaches a maximum iteration limit (set to five in this work).

\subsubsection{Semantic Scaffolding for Exemplar Generation}

Once a concrete concept is selected, this module constructs a semantic scaffold to guide the progressive generation of related visual content, as shown in the Semantic Exploration module in Figure~\ref{fig:pipeline}. The system initially builds a structured knowledge representation from lexical and commonsense resources before leveraging three analytical perspectives to drive the generation of coherent image exemplars.  

\paragraph{Structuring Multidimensional Semantic Relations}

To build a rich conceptual foundation, we query multiple knowledge bases, specifically ConceptNet 5~\cite{10.5555/3298023.3298212}, WordNet 3.0\footnote{WordNet [version 3.0]: https://wordnet.princeton.edu/}, and Wikidata\footnote{The Wikidata Query Service: https://query.wikidata.org/sparql}, to retrieve a set of relations associated with the input concept. 
We filter and organize these relations into three distinct semantic dimensions to guide the downstream generative process~\cite{10.1007/BF01263048, sowa1999knowledge, WINSTON1987417, 10.5555/176321.176324}:

\begin{itemize}[leftmargin=*, noitemsep]
    \item Taxonomic: This dimension includes hierarchical relations such as \textit{Hypernyms}, \textit{Hyponyms}, \textit{Synonyms}, \textit{Kind-of}, and \textit{Instance-of}.
    These relations clarify the concept's identity by situating it within a formal hierarchy, answering the question, ``What kind of thing is this?'' 
    \item Constitutive: This dimension focuses on partitive relations, including \textit{Part-Of} (meronymy) and \textit{Attribute-Of}.
    These relations break down the concept into its fundamental components and intrinsic properties, answering the question, ``What is this concept made of?''
    \item Associative: This dimension explores the concept's external connections, including its function (\textit{Used-for}), location (\textit{At-location}), and thematic relationships (\textit{Related-to}, \textit{Symbol-of}, \textit{Similar-to}). 
    These relations map how the concept functions in a broader context, answering the question, ``How is this concept used and what is it related to?''
\end{itemize}

\paragraph{Generating Exemplars via the Semantic Scaffold}

Instead of presenting the user with a raw graph of concept relations, the \sysname\ interface organizes this information into three explorable perspectives, each corresponding to one of the semantic dimensions:

\begin{itemize}[leftmargin=*, noitemsep]
    \item The Comparative View operationalizes the Taxonomic dimension. It displays the concept within its hierarchy, showing its parent classes (generalizations), sibling classes (alternatives), and subclasses (specializations). This view is designed to help users understand the concept itself more thoroughly by comparing it to similar entities.
    \item The Microscopic View operationalizes the Constitutive dimension. It offers a component-based diagram of the concept, showing its constituent parts and a detailed list of its intrinsic attributes. This view is designed to help users explore the internal details of the concept, enriching its visual features.
    \item The Macroscopic View operationalizes the Associative dimension. It places the concept at the center of a web, showing its functional, spatial, and other contextual links to related concepts. This view is designed to help users expand the scope of the concept by connecting it to external ideas and contexts.
\end{itemize}

Guided by this scaffold, the system generates a sequence of stylistically consistent exemplars using a chained, image-conditioned workflow.
An initial Comparative exemplar conditions the subsequent Microscopic generation, which in turn conditions the final Macroscopic output. 
We use the \texttt{FLUX.1-Kontext-dev} model~\cite{labs2025flux1kontextflowmatching} for this task.
This chained approach propagates stylistic elements across the steps, producing a sequence that is semantically progressive yet visually coherent.

\begin{figure}[b]
  \centering
  \includegraphics[width=\columnwidth]{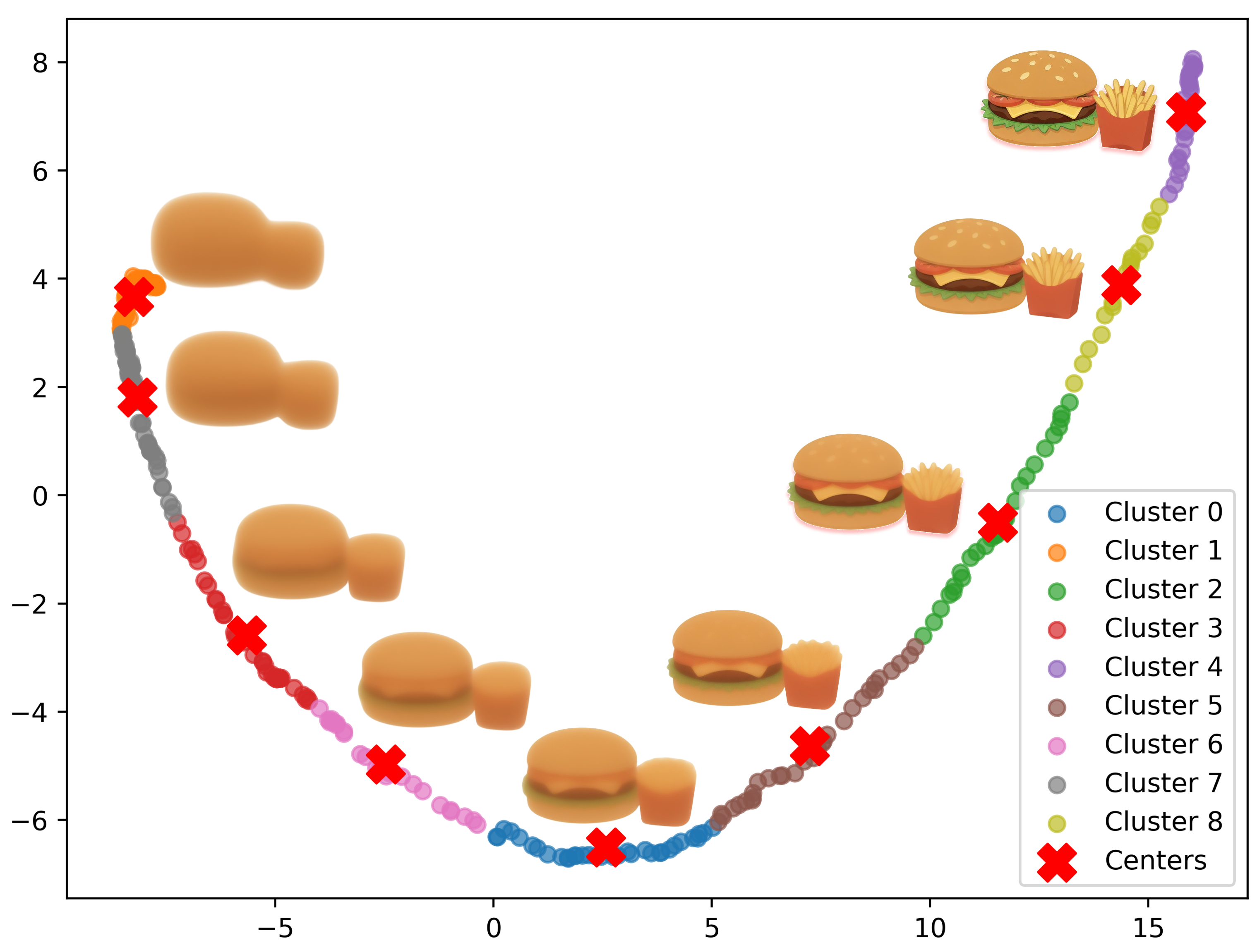}
  \caption{A series of simplified images clustered into nine groups and visualized in 2D via PCA. Red crosses denote cluster centers, highlighting the progression of simplification.}
  \Description{The figure shows a two-dimensional PCA scatterplot of hamburger-related images, with points colored into nine clusters. Along the curved distribution of points, large red X symbols mark the cluster centers. Above each center, a representative image is displayed, ranging from simplified blob-like burger shapes on the left to more detailed hamburgers with fries on the right.}
  \label{fig:k-means}
\end{figure}

\subsubsection{Progressive Simplification of Image Exemplars}

After generating a set of semantically varied exemplars, the third module of \sysname\ performs a visual simplification process on each one, shown in Figure~\ref{fig:pipeline}. 
This module is responsible for creating the progressive simplification sequence for each concept, which is a core component of our system. The process involves three main stages: 

\paragraph{SDS-Based Visual Simplification}
To transform a detailed exemplar into a sequence of progressively more abstract representations, we adopt a method based on Score Distillation Sampling (SDS)~\cite{Hertz_2023_ICCV, Liang_2024_CVPR} and Classifier-Free Guidance (CFG)~\cite{ho2022classifierfreediffusionguidance}. This approach leverages the feature-averaging effect of SDS to gradually remove fine details while preserving the overall macro-structural composition of the image. To mitigate potential shape distortion during this process, we increase the proportion of unconditional noise~\cite{11094816}. The simplification process is terminated based on a two-part criterion: first, we monitor the LPIPS score between images at fixed intervals, and when the score stabilizes, we use the Segment Anything Model (SAM)~\cite{ravi2024sam2segmentimages} to check if the resulting image consists of a single connected component. The process stops once this condition is met, ensuring a minimally abstract form.

\paragraph{K-means Clustering of Simplified Images}
The SDS process generates a long sequence of images. To select a discrete, representative set for the user interface, we perform k-means clustering. We first extract deep features from each image in the sequence using an ImageNet-pretrained ResNet-18~\cite{He_2016_CVPR}. 
We then perform k-means clustering in this feature space (with $k=9$ in our implementation, see an example in Figure~\ref{fig:k-means}). Using a nearest-centroid strategy, we select the image closest to each cluster's center as its representative. This approach ensures that the selected images are both diverse (covering different visual patterns from the sequence) and representative (typical of each stage of simplification).

\paragraph{Semantic Segmentation and Layering with SAM} 
In the final stage, we reapply SAM~\cite{ravi2024sam2segmentimages} to the representative images selected from the clustering step to extract their corresponding object masks. 
These masks are then layered to create the final icon representations. 
Each mask is assigned partial transparency ($Alpha = 0.5$ in our work) and ordered by area, with larger masks placed in the back layers and the original simplified image at the bottom as a background. 
This layered strategy preserves the precise contours of the objects via the masks while retaining fine details from the underlying simplified image, resulting in a set of icons that maintain a clear overall form while exhibiting varying levels of visual detail.

\subsubsection{Stylization and Refinement of Icon Representations}

The layered masks produced by the previous abstraction module, while structurally correct, often contain visual noise and irregular artifacts. Therefore, the final module of our pipeline is an icon refinement process designed to clean these outputs and generate icons in several consistent styles, addressing the need for stylistic cohesion, as shown in Figure~\ref{fig:pipeline}.

To implement this, we fine-tune the \texttt{FLUX.1-Kontext-dev} model using Low-Rank Adaptation (LoRA) on a curated dataset comprising three common visual variants: Outline, Filled, and Color.

First, for initial refinement and noise removal, we trained a \texttt{Mask-to-Outline} LoRA. The training data consisted of approximately 20–30 pairs of SAM-segmented images and their corresponding clean, original icons from our collected dataset. 
Applying this model to the noisy masks from the previous stage transforms them into standardized stroke-based icons.
Second, to generate the remaining visual variants, we fine-tuned two additional LoRA models (i.e., \texttt{Outline-to-Filled} and \texttt{Outline-to-Color}) on small datasets ($\sim$20–30 pairs each).
These models transform the standardized outline icon into its corresponding filled and colored counterparts.
This multi-stage refinement process yields the final icon grid, where each representation is available in three distinct and stylistically consistent visual variants, ready for presentation in the user interface.

\subsection{User Scenario}

To illustrate how \sysname\ streamlines the workflow, we present a scenario involving Tom, a designer tasked with developing a cohesive icon suite for a new ``fast food'' delivery application. He initiates the process by inputting the target concept directly.

Tom begins by entering ``fast food'' into the Input Concept module (Figure~\ref{fig:interface}.A). The system analyzes the concept and presents a candidate table with related options, such as ``Hamburgers'', ``Cheese Slice'', and ``French Fry'' (Figure~\ref{fig:interface}.B). 
Each candidate is accompanied by quantitative ratings for Concreteness, Familiarity, Imageability, and Meaningfulness. This allows Tom to quickly assess which concepts are most promising. After reviewing the results, he selects ``Hamburgers'' as the core concept for the initial generation.

Next, Tom moves to the Semantic Exploration module to broaden his creative scope (Figure~\ref{fig:interface}.C). 
A tree-structure visualization shows him the semantic associations for ``Hamburgers'', organized by the three analytical perspectives: Taxonomic, Constitutive, and Associative (Figure~\ref{fig:interface}.C1). 
Using the stepwise semantic progression prompts (Figure~\ref{fig:interface}.C2), Tom guides the system to generate a distinct, stylistically consistent image exemplar for each of the Comparative, Microscopic, and Macroscopic views. This process yields a series of semantically varied images, from which he selects his preferred outputs to serve as the foundation for the icon grid.

Tom then proceeds to the Visual Simplification module (Figure~\ref{fig:interface}.D), where the system automatically generates a progressive simplification sequence for each selected exemplar, transforming the detailed inputs into a spectrum of simplified versions (Figure~\ref{fig:interface}.D1). The complete results are organized into the dual-axis exploration canvas (Figure~\ref{fig:interface}.E1), a two-dimensional grid structured by visual complexity (from detailed to abstract) and semantic richness (from simple to elaborate). This grid allows Tom to efficiently compare and evaluate the design trade-offs.

While exploring the grid, Tom notices that some of the more complex icons are not clearly distinguishable at a small scale. He revisits the visual representation module (Figure~\ref{fig:interface}.D1 \& D2), selects a few different simplified versions, and regenerates the grid. 
This iterative loop allows him to quickly refine his selections until he is satisfied with the final set of outline-style icons. 
As a final step, he uses the style refinement feature to generate filled-style variants (Figure~\ref{fig:interface}.E2), further expanding the utility of his design results for different use cases. This scenario demonstrates how \sysname\ effectively supports designers in concept analysis, semantic exploration, and visual representation, facilitating an iterative and multidimensional icon creation process.

\section{Hypotheses}

Prior research highlights the inherent tension in icon design between maintaining visual consistency and exploring diverse semantic options~\cite{10.1145/3313831.3376618, 10.1145/3618364, 10.1145/3664647.3681057, 10.1145/3531065}.
While HCI literature suggests that exploring multiple design alternatives in parallel promotes divergent thinking and higher-quality outcomes~\cite{10.1145/1879831.1879836}, doing so manually imposes high extraneous cognitive load, which can conversely inhibit creativity~\cite{malycha2017enhancing}.
\sysname\ is designed to resolve this conflict by automating the progressive simplification (C2) and decoupling semantic control from stylistic rendering (C1, C3). 
By offloading the execution burden while structuring the design alternatives into a navigable icon grid (C2), we expect \sysname{} to support rapid exploration and ideation without the associated cognitive cost. 
Grounded in these specific affordances, we propose the following hypotheses relative to a baseline workflow:

\begin{enumerate}[leftmargin=*, noitemsep, label={\textbf{H\arabic*}}]
    \item Compared to the baseline, \sysname\ will enhance the icon grid design workflow, resulting in a higher perceived usability and a better overall user experience.
    \item Compared to the baseline, 
    \sysname\ will significantly reduce designers' cognitive load in the task of generating icon grids with varying levels of complexity by: (H2a) decreasing mental demand, (H2b) minimizing physical demand, (H2c) reducing perceived time pressure, (H2d) increasing satisfaction with their performance, (H2e) lowering the overall effort required, and (H2f) reducing frustration.
    \item Compared to the baseline, \sysname\ will provide superior creativity support, enabling designers to: (H3a) more effectively explore a broad range of creative ideas, (H3b) experience a more effective human-AI collaboration, (H3c) feel more enjoyment and engagement in the design process, (H3d) perceive the results as more valuable for the effort expended, (H3e) maintain a higher level of immersion in the task, and (H3f) achieve greater creative expressiveness.
    \item Compared to the baseline, \sysname\ will demonstrate superior performance in progressive icon grid design, resulting in higher-quality outcomes and greater user satisfaction. Specifically, \sysname\ will produce outcomes characterized by: (H4a) a clearer progression of semantic richness, (H4b) more distinct gradients of visual complexity, (H4c) higher stylistic consistency, (H4d) better alignment with intended concepts, (H4e) more creative design outcomes, (H4f) a wider diversity of solutions, and (H4g) higher overall satisfaction with the final product.
\end{enumerate}

\section{Evaluation}

\subsection{Participants}

We recruited 32 participants (14 female, 18 male; ages 20–26, $M = 22.6, SD = 1.54$) from a university-affiliated graphics group and online design communities.
This strategy targeted individuals with a demonstrated interest in visual creation, ensuring a shared baseline of visual literacy.
The sample comprised 20 participants with computer science backgrounds and 12 with design backgrounds. 
Reflecting the interdisciplinary nature of human-AI co-creation for visual tasks (e.g., ~\cite{doi:10.1126/science.adh4451}), we viewed this distribution not as a binary split, but rather as representative of the technical and artistic intersection required for modern generative workflows.
This sampling strategy allowed us to evaluate \sysname's robustness across the diverse range of skills found in real-world creative support tool usage~\cite{10.1145/3290605.3300619}.
All participants reported prior experience with AI-based image generation tools. Participants received \$8 USD as compensation for a session lasting approximately one hour.

\begin{figure*}[t]
  \centering
  \includegraphics[width=\textwidth]{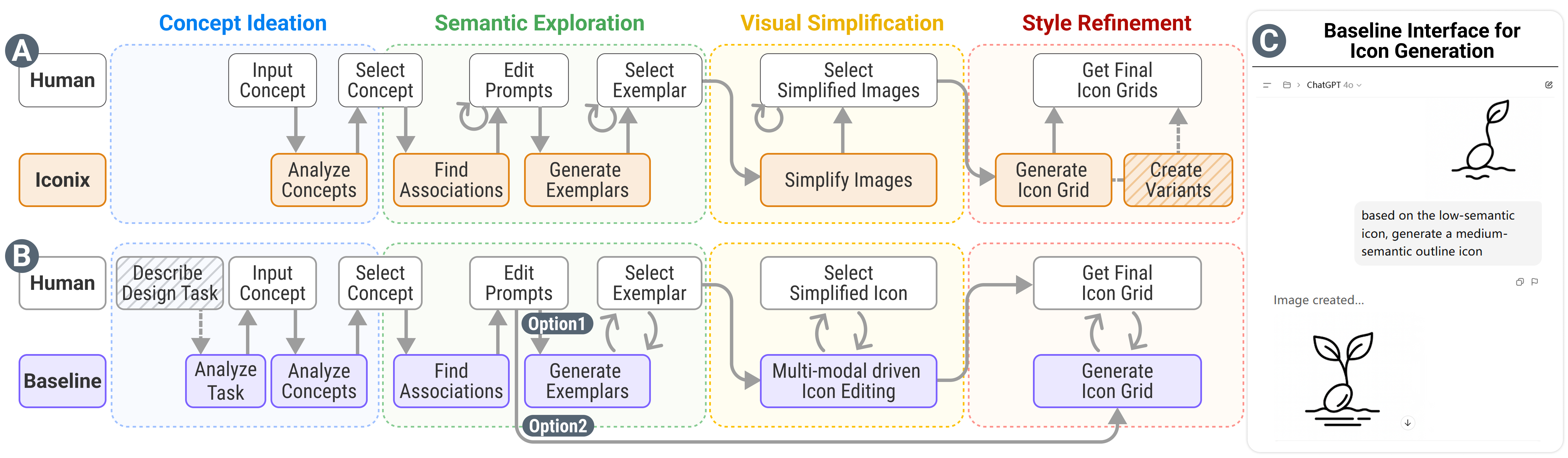}
  \caption{Workflow comparison between (A) \sysname\ and (B) the Baseline. Both systems share the same major steps for icon generation. The key difference is the linear, selection-based progression of \sysname\ compared to the cyclic, prompt-based iteration required in the Baseline. Curved lines indicate iterative operations, while modules with diagonal hatching indicate optional actions. (C) is the Baseline (i.e., ChatGPT) interface used during the study.
  }
  \label{fig:baseline}
  \Description{The figure presents a side-by-side workflow comparison between Iconix and the Baseline system. The upper workflow (A) shows the Iconix process arranged in a mostly linear sequence across four stages: concept ideation, semantic exploration, visual simplification, and style refinement. Each stage contains labeled modules connected by straight arrows, with occasional curved arrows indicating optional iterative adjustments. The lower workflow (B) shows the Baseline process, which includes additional user-provided task descriptions and more frequent looping arrows that represent repeated prompt editing and exemplar generation. Some blocks contain diagonal hatching to indicate optional steps. On the right, panel C displays a screenshot of the Baseline chat interface, showing a user prompt requesting a medium-semantic outline icon and the system’s generated sprout illustration.}
\end{figure*}

\subsection{Apparatus and Conditions}

\begin{figure*}[b]
  \centering
  \includegraphics[width=\textwidth]{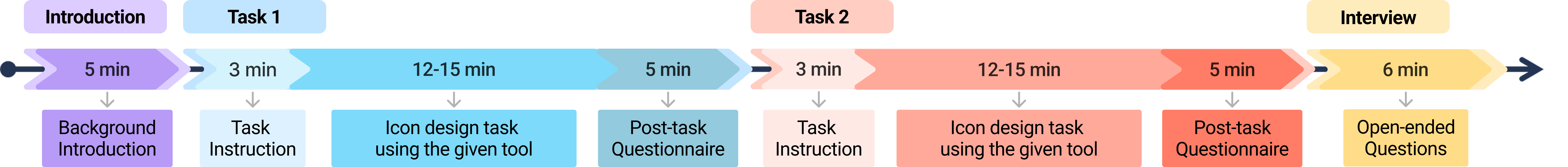}
  \caption{The user study procedure. The study followed a within-subjects design where participants completed two design tasks using both \sysname\ and the Baseline, with the order of conditions and tasks counterbalanced to ensure fairness.
  }
  \label{fig:study_proceduce}
  \Description{The figure shows a horizontal timeline illustrating the sequence of activities in the user study. It begins with a 5-minute Introduction segment that includes a background briefing. Task 1 follows, consisting of a 3-minute instruction period, a 12 to 15-minute icon-design activity using the assigned tool, and a 5-minute post-task questionnaire. Task 2 mirrors the same structure with its own instruction, design session, and questionnaire. The timeline concludes with a 6-minute interview segment featuring open-ended questions.}
\end{figure*}

We compared \sysname\ against ChatGPT\footnote{ChatGPT: https://chatgpt.com/} to isolate the benefits of our proposed AI-supported visual ideation and abstraction controls relative to a generalist baseline. 
We prioritized this comparison over manual tools (e.g., Adobe Illustrator) to avoid proficiency confounds, focusing specifically on generative support for early-stage ideation. 
Additionally, given its accessibility and zero-setup cost, ChatGPT serves as an ecological baseline representing the current\footnote{Refers to the commercially available version of ChatGPT (powered by GPT-4o) during the study period (June--August 2025).} multi-modal generation workflow available to general users who approach creative tasks~\cite{10.1145/3706598.3713259}.
Crucially, this comparison isolates the value of \sysname's domain-specific scaffolding against the flexible but often unguided nature of standard generative AI tools, enabling us to investigate how structured workflows preserve human design intent and agency compared to raw machine intelligence.
Figure~\ref{fig:baseline} presents a detailed comparison of the user actions and system responses for both \sysname\ and the Baseline.
The study employed a within-subjects design where each participant completed tasks under two conditions:
\begin{itemize}[leftmargin=*, noitemsep]
    \item Baseline Condition: Participants used the standard ChatGPT web interface backed by GPT-4o. They leveraged its multi-modal capabilities for text-to-image generation and editing to create a complete icon grid that matched the same design brief used in the \sysname{} condition.
    \item \sysname\ Condition: Participants used \sysname, our custom-developed, web-based system. The system provided the structured semantic scaffolding and visual abstraction features described in Section~\ref{sec:system}.
\end{itemize}

The order of conditions was counterbalanced across participants using a Latin square design to mitigate learning and order effects. We conducted the experiment in a controlled laboratory setting, with each participant using a Windows 11 desktop computer equipped with a 27-inch 4K display and the latest version of Google Chrome.

\subsection{Tasks and Procedure}

We designed two tasks to evaluate the systems' performance on concepts that differ in the specificity of their visual imagery.
The first was a concrete task (T1) using the concept ``Hamburger''. This concept was chosen because its visual imagery is more specific and widely understood, requiring less metaphorical interpretation from the designer.
The second was an abstract task (T2) using the concept ``Hope''. This concept was chosen because it lacks a direct, specific visual counterpart and requires the designer to explore more metaphorical or symbolic imagery. For this task, participants were permitted to use the direct concept or a concrete proxy they felt was representative.

Each participant completed both tasks, one in each of the two conditions. 

The procedure for each session (shown in Figure~\ref{fig:study_proceduce}) was as follows:
\begin{enumerate}[leftmargin=*, noitemsep]
    \item Briefing and Training: Participants were briefed on the study goals and the core concepts of semantic richness and visual complexity, then provided a standardized seed prompt to align initial conditions and mitigate the cold-start problems (see Appendix D).
    \item First Condition: Participants were assigned their first condition (either Baseline or \sysname) and corresponding task (T1 or T2). They were asked to create a set of icons for the given concept using the assigned system.
    \item Post-Task Questionnaire: After completing the task, participants filled out a questionnaire assessing their experience with the system.
    \item Second Condition: Participants then moved to the second condition, using the other system to complete the remaining task.
    \item Second Post-Task Questionnaire: Participants completed the same questionnaire for the second condition.
    \item Semi-Structured Interview: At the end of the session, we conducted a brief semi-structured interview to gather qualitative feedback, including suggestions for system improvement and reflections on their creative process.
\end{enumerate}

\subsection{Data Collection}

We employed a mixed-methods approach to evaluate our hypotheses, collecting quantitative data from questionnaires and qualitative data from design artifact analysis and post-session interviews.

\paragraph{Quantitative Measures}
Following each of the two task conditions, participants completed a post-task questionnaire.
This included an adaptation of standard validated scales and custom-designed 7-point Likert-scale questions (see Appendix A).
A summary of all measured factors is presented in Table~\ref{tbl:result}.

\begin{itemize}[leftmargin=*, noitemsep]
    \item System Usability Scale (SUS): We utilized the standard 10-item SUS questionnaire~\cite{brooke1996sus} to assess participants' subjective perceptions of system usability.
    \item NASA Task Load Index (NASA-TLX): We adopted the NASA-TLX~\cite{HART1988139} to assess perceived cognitive workload while performing icon grid design tasks.
    \item Creativity Support Index (CSI): To measure how well each system supported creative work, we used the CSI~\cite{10.1145/2617588}, adapting its items to the specific context of icon grid design with generative AI.
    \item Custom Questionnaire: We developed a custom questionnaire using 7-point Likert-scale items to assess simple, unidimensional constructs. These items measured participants' satisfaction with their final designs and their perception of the system's support for semantic and style controls.
\end{itemize}

\paragraph{Qualitative Measures} 

We collected two forms of qualitative data: the final design artifacts and post-session semi-structured interviews.
\begin{itemize}[leftmargin=*, noitemsep] 
    \item Design Artifacts: All icon grids created by participants during both task conditions were collected for qualitative review.
    \item Semi-Structured Interviews: The interview questions were designed to elicit comparisons between the two conditions, exploring the design workflow, challenges encountered, and preferences (see Appendix C). With participants' consent, these interviews were audio-recorded and later transcribed for analysis.
\end{itemize}

\section{Results Analysis}

\begin{table*}[t]
\centering
\caption{Statistical comparison of user feedback for \sysname\ and the Baseline. The table includes descriptive statistics (Mean, Standard Deviation), inferential test results ($t, Z$), effect sizes (Cohen's $d, r$), unadjusted p-values ($p$), and Holm-Bonferroni adjusted p-values ($p_{\text{holm}}$) to control for multiple comparisons.
Significance ($-$: $p>.100$, $+$: $.050<p<.100$, $*$: $p<.050$, $**$: $p<.010$, $***$: $p<.001$) is based on $p_{\text{holm}}$, except for the \textit{Usability} construct, which uses the original $p$-value. Underlined statistics denote non-parametric results from the Wilcoxon analysis.
Arrows indicate preferred outcomes ($\downarrow$: lower is better; $\uparrow$: higher is better).
}

\begin{tabular}{llrrrrrrrrr}
\toprule
\multicolumn{1}{c}{\multirow{2}{*}{\textbf{Construct}}} &
\multicolumn{1}{c}{\multirow{2}{*}{\textbf{Factor}}}   &
\multicolumn{2}{c}{\sysname} & 
\multicolumn{2}{c}{Baseline} & 
\multicolumn{4}{c}{Statistics} & 
\multicolumn{1}{c}{\multirow{2}{*}{\textbf{Hypothesis}}} \\
\cmidrule(lr){3-4}\cmidrule(lr){5-6}\cmidrule(lr){7-10}
         &                   & Mean            & SD    & Mean     & SD      & $t/\underline{Z}$                & $d/\underline{r}$                & $p$                & $p_{\text{holm}}$ &  \\
\toprule
\textbf{Usability}~\cite{brooke1996sus} & 
Perceived Usability~$\uparrow$ 
& 79.531              & 10.915 & 63.594   & 12.426             & 5.759           & 1.018   &  \textless{}$.001^{***}$           & N/A    &  \textbf{H1} \textbf{Accepted}          \\
\midrule
\multicolumn{1}{l}{\multirow{6}{*}{\parbox{1.6cm}{\textbf{Cognitive Load}~\cite{HART1988139}}}} 
                 & Mental Demand~$\downarrow$       & 2.406    & 1.341 & 3.406     & 1.292    & \underline{2.747}     & \underline{.614} & .006 & $.006^{**}$     &   \textbf{H2a} \textbf{Accepted}   \\
                 & Physical Demand~$\downarrow$               & 1.625            & .871 & 2.750    & 1.606    & \underline{3.444} & \underline{.835}          & \textless{}.001            & $.002^{**}$   &   \textbf{H2b} \textbf{Accepted}         \\
                 & Temporal Demand~$\downarrow$        & 3.594       & 1.720 & 4.688     & 1.469          & \underline{3.034} & \underline{.583}       & .002     & $.005^{**}$    &  \textbf{H2c} \textbf{Accepted}          \\
                 & Performance~$\uparrow$          & 5.313      & .859 & 4.313     &1.176             & \underline{3.050}           & \underline{.665} & .002        & $.007^{**}$    &  \textbf{H2d} \textbf{Accepted}          \\
                 & Effort~$\downarrow$        & 2.844        & 1.247 & 4.188     & 1.230              & 4.431            & .783 & \textless{}.001      & \textless{}$.001^{***}$    &  \textbf{H2e} \textbf{Accepted}          \\
                 & Frustration~$\downarrow$             & 2.094       & .963 & 3.438     & 1.585       & 4.481      & .792 & \textless{}.001       & \textless{}$.001^{***}$    &  \textbf{H2f} \textbf{Accepted}          \\

\midrule
\multicolumn{1}{l}{\multirow{6}{*}{\parbox{1.7cm}{\textbf{Creativity Support}~\cite{10.1145/2617588}}}}       & Exploration~$\downarrow$            & 2.406            & 1.500  & 3.688     & 1.749
              & 3.334           & .589 & .002            & $.007^{**}$   &   \textbf{H3a} \textbf{Accepted}         \\
                 & Collaboration~$\downarrow$           & 2.156              & 1.370  & 3.531     & 1.586              & 4.065            & .719 & \textless{}.001     & $.002^{**}$   &   \textbf{H3b} \textbf{Accepted}         \\
                 & Engagement~$\downarrow$              & 2.438            & 1.413 & 3.625     & 1.621        & 3.490      & .617 & .001         & $.006^{**}$   &   \textbf{H3c} \textbf{Accepted}         \\
                 & Results Worth Effort~$\downarrow$ & 2.094            & 1.254 & 3.188     & 1.355       & 3.933      & .695 & \textless{}.001   & $.002^{**}$    &  \textbf{H3d} \textbf{Accepted}          \\
                 & Immersion~$\downarrow$      & 2.656      & 1.537 & 3.094     & 1.400        & \underline{1.540}    & \underline{.314} & .124            & $.124^{-}$    &   \textbf{H3e} Rejected         \\
                 & Expressiveness~$\downarrow$     & 2.938        & 1.605 & 3.469     & 1.685               & 1.641    & .290 & .111            & $.222^{-}$    &   \textbf{H3f} Rejected         \\

\midrule
\multicolumn{1}{l}{\multirow{7}{*}{\parbox{1.62cm}{\textbf{Design\\Satisfaction}}}}
            & Semantic Richness~$\uparrow$          & 5.844           & 1.247 & 3.750       & 1.646              & 5.559           & .983 & \textless{}.001 & \textless{}$.001^{***}$  &  \textbf{H4a} \textbf{Accepted}          \\
            & Concept Faithfulness~$\uparrow$          & 5.625        & 1.157 & 4.906     & 1.329         & \underline{2.436}        & \underline{.497} & .015 & $.015^{*}$  &  \textbf{H4b} \textbf{Accepted}          \\   
            & Visual Complexity~$\uparrow$          & 5.594        & 1.103 & 3.938     & 1.625         & 4.903           & .867 & \textless{}.001 & \textless{}$.001^{***}$  &  \textbf{H4c} \textbf{Accepted}          \\
            & Style Consistency~$\uparrow$          & 6.094       & .963 & 4.938    & 1.458         & 3.877           & .685 & \textless{}.001 & $.001^{**}$  &  \textbf{H4d} \textbf{Accepted}          \\
            & Outcome Creativity~$\uparrow$                  & 5.719       & 1.023 & 4.438   & 1.435        & \underline{3.770}      & \underline{.739} & \textless{}.001   & \textless{}$.001^{***}$    &   \textbf{H4e} \textbf{Accepted}         \\
            & Outcome Diversity~$\uparrow$          & 5.531          & 1.191 & 4.094      & 1.467        & 5.665       & 1.001 & \textless{}.001 & \textless{}$.001^{***}$  &  \textbf{H4f} \textbf{Accepted}          \\
             & Overall Satisfaction~$\uparrow$                 & 5.813    & .896 & 4.531    & 1.244      & \underline{3.792}     & \underline{.774} & \textless{}.001    & \textless{}$.001^{***}$    &  \textbf{H4g} \textbf{Accepted}          \\
\bottomrule
\end{tabular}
\label{tbl:result}
\Description{Statistical comparison of user feedback between Iconix and the Baseline. The table presents descriptive statistics (Mean, SD) alongside inferential metrics, including test statistics (t, Z), effect sizes (Cohen’s d, r), and p-values. Significance is determined using Holm-Bonferroni adjusted p-values (p_holm) to control for multiple comparisons, with the exception of the Usability construct, which relies on unadjusted values. Underlined figures denote non-parametric results from the Wilcoxon analysis. Arrows indicate the preferred direction of the outcome ($\downarrow$ denotes lower is better; $\uparrow$ denotes higher is better).
}
\end{table*}

This section presents the statistical analysis of participant ratings from our within-subject experiment. 
All quantitative results, including descriptive statistics for both conditions, inferential statistics, and effect sizes, are detailed in Table~\ref{tbl:result}.
To ensure statistical rigor, we applied a Holm-Bonferroni correction~\cite{29def780-e117-38f0-8afb-edf384af3fad} to control the family-wise error rate from multiple comparisons.
We therefore report both the original p-value ($p$) and the adjusted p-value ($p_{\text{holm}}$).
Notably, for non-parametric tests, the effect size $r$ was calculated using $r = Z / \sqrt{N_{\text{effective}}}$, where $N_{\text{effective}}$ represents the number of pairs with non-zero differences.
In the following sections, these findings are discussed for each evaluation criterion, contextualized with participants' qualitative feedback.

\subsection{System Usability}

We verified all relevant statistical assumptions prior to analysis to ensure the validity of our findings.
A Shapiro-Wilk test confirmed the normality of the participants' ratings ($W(32) = .959, p = .250$), justifying the use of a parametric test.
A paired-samples t-test revealed that participants rated \sysname\ significantly higher in usability than the baseline.
As detailed in Table~\ref{tbl:result}, this difference was not only statistically significant but also represented a large effect size,
indicating a substantial improvement in perceived usability. 
These findings confirm \textbf{H1}.
Qualitative feedback from the post-session interviews helps explain this significant difference in usability scores. 
The majority of participants (30 out of 32) described \sysname\ as easy to learn and use. 
Two participants ({P12, Male, 24; P22, Female, 23}) initially found the interface ``\textit{more information-dense}'' than the baseline's conversational format, but both concluded it was ``\textit{quite effective overall}'' after a brief period of familiarization.

Participants attributed the improved experience primarily to the design of the exploration canvas. Unlike the linear, conversational nature of the baseline, which required them to scroll back to compare previous outputs, \sysname's matrix-style layout (Figure~\ref{fig:interface}.E) presented all generated icons simultaneously. 
This interface enabled them to access and compare multiple design alternatives at a glance, which, as several participants noted, significantly improved their workflow by eliminating the need to recall or navigate back through earlier outputs.

\subsection{Cognitive Load}

We assessed cognitive load using the six dimensions of the NASA-TLX: Mental Demand (the mental effort required), Physical Demand (the degree of physical exertion), Temporal Demand (time-related pressure), Performance (self-assessed quality of task completion), Effort (the overall amount required), and Frustration (dissatisfaction experienced).
We tested all six dimensions for normality using Shapiro-Wilk tests. This check revealed that the data for mental demand, physical demand, temporal demand, and performance were not normally distributed (all $p <.05$). Conversely, the data for effort and frustration were normally distributed. Accordingly, we applied Wilcoxon signed-rank tests for the four non-normal dimensions and paired-samples t-tests for the remaining two.

\begin{figure}[t]
  \centering
  \includegraphics[width=0.92\columnwidth]{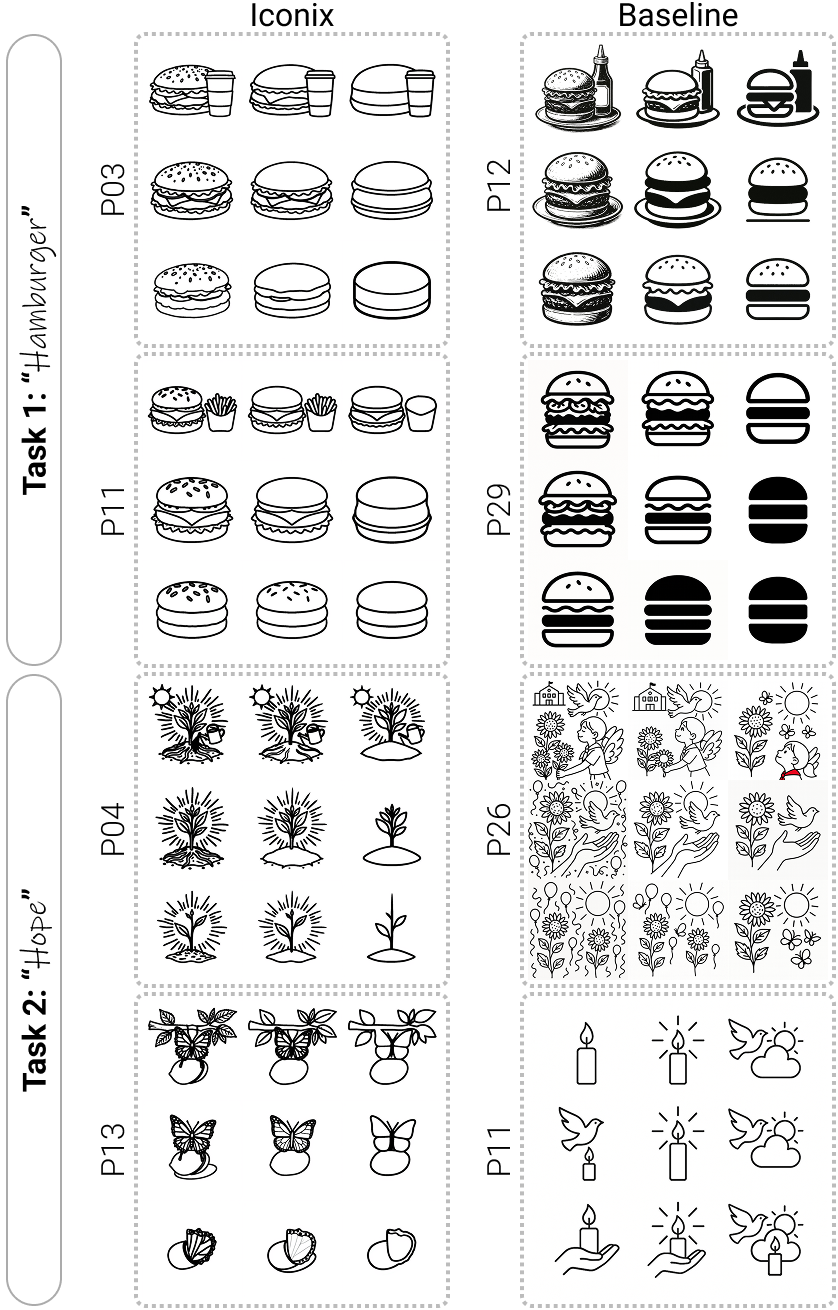}
  \caption{Eight sample icon grid outputs generated by the \sysname\ and the Baseline for topics of T1 and T2.}
  \Description{The figure displays eight icon-grid examples arranged in two rows. Under Task 1 “Hamburger,” four grids show different sets of hamburger-related icons, including detailed outline drawings with fries, simplified line-based burgers, filled-shape silhouettes, and progressively abstracted burger forms. Under Task 2 “Hope,” four grids depict symbolic representations such as hands holding flowers, sun and plant motifs, candles with rays of light, doves, and butterflies. Each grid contains multiple variations of the concept rendered in consistent styles and levels of visual complexity.}
  \label{fig:user-study}
\end{figure}

In support of \textbf{H2a}–\textbf{H2f}, our analysis found that \sysname\ significantly reduced cognitive load across all six dimensions compared to the baseline.
As detailed in Table~\ref{tbl:result}, these differences all favored \sysname\ and remained statistically significant after a Holm-Bonferroni correction. 
The medium to large effect sizes indicate that \sysname\ provided a substantial practical reduction in cognitive burden compared to the baseline.
Qualitative feedback from our interviews explains this significant reduction.
Participants reported that achieving satisfactory outcomes with the baseline required substantial effort due to the need for ``\textit{continuous reflection and repeated, step-by-step image editing operations}''. For instance, one participant (P26, Female, 22) described her manual process of simplifying an icon for ``hope'' in the baseline, which involved a series of discrete edits: ``\texttt{removing the bird and the hand, adding flowers}'', then ``\texttt{removing the lines and adding two butterflies}'', and finally ``\texttt{removing the flowers, balloons, and lines}'' (see P26 in Figure~\ref{fig:user-study}). She described the experience as mentally taxing, stating, ``\textit{it felt like I had to do all the heavy lifting in my head for every single change}''.
In contrast, participants found that \sysname\ streamlined this process, enabling them to achieve desired results with greater ease. Several participants (P16, Male, 23; P32, Female, 21) specifically highlighted the value of the semantic exploration module, noting that the tree-structure visualization ``\textit{helped reduce (their) cognitive burden}'' during the initial ideation phase.

\subsection{Creativity Support}

We also assessed creativity support using our adapted CSI questionnaire, which measured six dimensions: Exploration (the perceived ease of exploring design options), Collaboration (the perceived effectiveness of the user-AI collaboration), Enjoyment (the level of enjoyment in the creative process), Results Worth Effort (the perceived worthwhileness of the results), Immersion (the degree of focus), and Expressiveness (the support for creative expression).
A check for normality was conducted on all six dimensions. This Shapiro-Wilk analysis revealed that the data for Immersion were not normally distributed ($W(32) = .929, p = .038$), whereas the remaining five dimensions were (all $p > .05$). Accordingly, we used a Wilcoxon signed-rank test to analyze Immersion and paired-samples t-tests for the other five dimensions.

As detailed in Table~\ref{tbl:result}, our analysis for creativity support showed mixed results. \sysname\ was rated as significantly outperforming the baseline on four of the six dimensions: Exploration, Collaboration, Enjoyment, and Results Worth Effort. 
These differences all represented large effect sizes and remained statistically significant after a Holm-Bonferroni correction. 
This suggests that \sysname\ excels at improving the process of creativity by facilitating easier design exploration (\textbf{H3a}), enabling more effective user-AI collaboration (\textbf{H3b}), and making the overall workflow more engaging (\textbf{H3c}) and rewarding (\textbf{H3d}).
Conversely, we found no significant differences for Immersion or Expressiveness.
This suggests that while \sysname\ streamlines the process of creativity, our analysis found no statistically significant difference between it and the baseline regarding the user's subjective sense of flow (\textbf{H3e}) or the perceived creative expressiveness of the final results (\textbf{H3f}).
These findings provide support for \textbf{H3a–d}, but not \textbf{H3e} and \textbf{H3f}.

Qualitative feedback from our interviews helps explain these findings. Participants felt that \sysname\ provided a significantly more effective starting point for ideation, which substantially enriched the co-creative experience by making the workflow more rewarding and less burdensome. 
P29 (Male, 22) contrasted the two experiences sharply: ``\textit{The baseline was a constant stop-and-start. I'd have an idea, but then I'd get bogged down in all the little edits, and by the time I was done, I'd lost my train of thought.}''
P29 felt \sysname\ was different because ``\textit{...it let me just focus on the ideas first... It felt like (\sysname) separated the `thinking' part from the `doing' part.}'' 
This separation of concerns made the process feel more streamlined and goal-oriented for participants, and the AI's role was seen as being more practically helpful.
However, the perceived improvement in the process did not, on average, translate into a statistically significant higher rating of subjective flow or creative expression in the final outcome.
This was the case even though a subset of participants (P4, Male, 23; P7, Male, 24; P29, Male, 22; P32, Female, 21) did individually report that \sysname\ allowed them to ``\textit{stay focused on the flow of ideas}'', a sentiment that was not reflected in the overall statistical trend.

\subsection{Design Satisfaction}

Finally, we assessed design satisfaction using our custom questionnaire, which measured seven factors: semantic richness (perceived progression in semantic range), concept faithfulness (fidelity in conveying the intended concept), visual complexity (perceived gradient in visual detail), style consistency (stylistic unity of the final set), outcome creativity (perceived creativity of the final designs), outcome diversity (perceived diversity of the final designs), and overall satisfaction (holistic satisfaction with the final outcome).
We confirmed the scale's reliability using Cronbach's alpha ($\alpha$), which demonstrated good internal consistency for both the \sysname\ ($\alpha = .823$) and baseline ($\alpha = .864$) conditions.
Shapiro-Wilk normality tests were performed on all seven factors. 
This analysis revealed that data for concept faithfulness ($W(32) =.889, p =.003$), outcome creativity ($W(32) =.925, p =.029$), and overall satisfaction ($W(32) =.922, p =.024$) were not normally distributed. 
As the remaining four factors were found to be normal (all $p > .05$), we applied Wilcoxon signed-rank tests for these three factors and paired-samples t-tests for the other four.

Overall, \sysname\ was rated significantly higher than the baseline across all seven measures, as detailed in Table~\ref{tbl:result}.
We observed this first in the core control mechanisms. 
Participants rated \sysname\ as providing superior control over semantic properties, including significantly clearer semantic progression (\textbf{H4a}) and higher ratings for concept faithfulness (\textbf{H4b}). 
This was reflected in style control, where \sysname\ also had a more distinct visual complexity gradient (\textbf{H4c}) and received significantly higher ratings for style consistency (\textbf{H4d}). 
These improvements in the design process ultimately led to a better final product, with participants reporting significantly higher outcomes in creativity (\textbf{H4e}), diversity (\textbf{H4f}), and overall satisfaction (\textbf{H4g}) for the designs created with \sysname.
All seven of these findings remained statistically significant after a Holm-Bonferroni correction and represented large to very large effect sizes, indicating a substantial and practical improvement in the entire design satisfaction.
Therefore, all these findings support \textbf{H4a–H4g}.

\begin{figure*}[t]
  \centering
  \includegraphics[width=\textwidth]{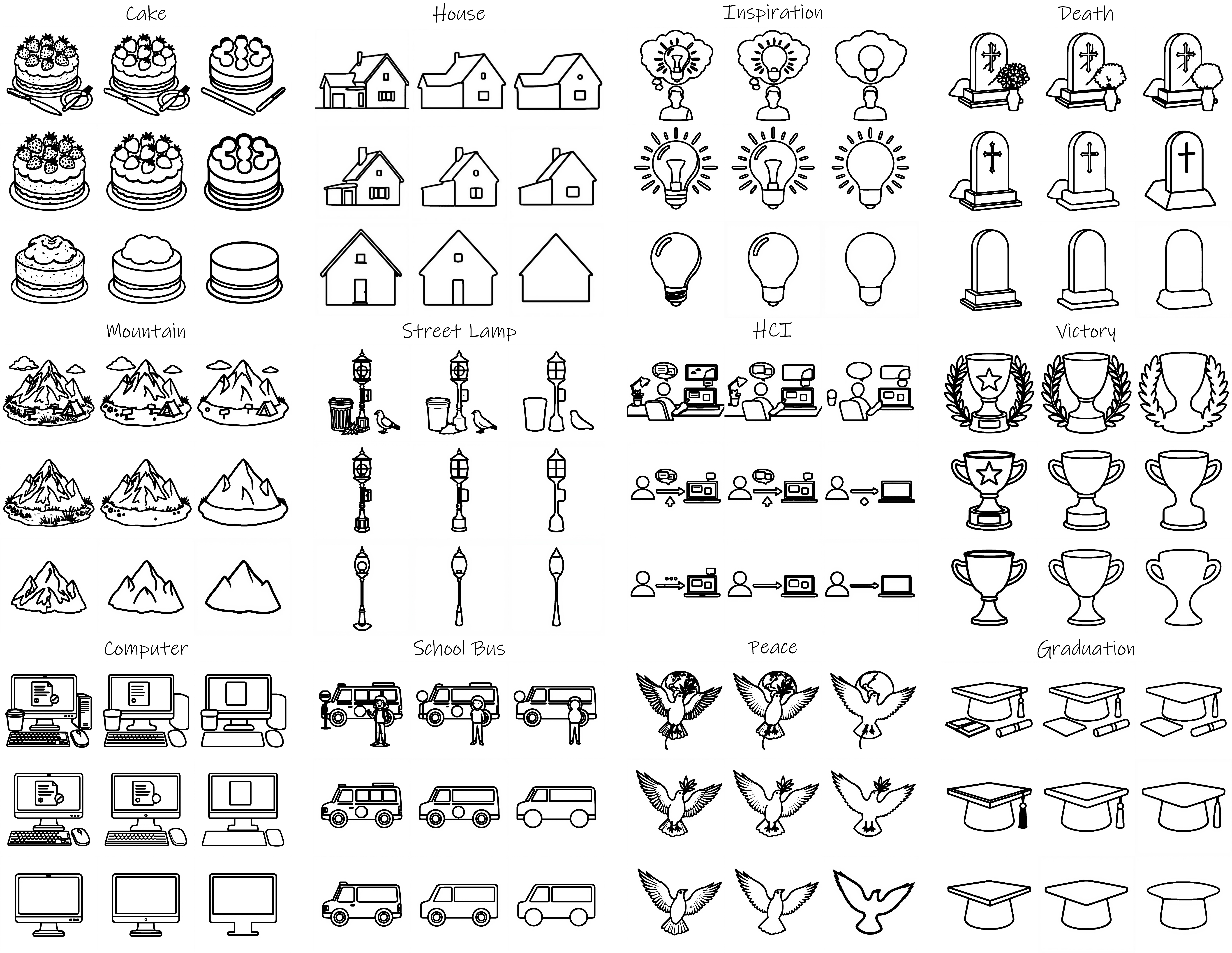}
  \caption{\sysname\ generates style-consistent icon grids of varying complexity from user-specified concepts. The figure contains 12 concepts: the six on the left are concrete, and the six on the right are abstract. For each concept, icons are organized along two orthogonal dimensions: from top to bottom, semantic richness decreases (from elaborate to simple); from left to right, visual complexity decreases (from detailed to abstract).}
  \label{fig:result}
  \Description{The figure consists of twelve icon grids arranged in three rows and four columns. The left six grids depict concrete concepts such as cakes, houses, gravestones, mountains, street lamps, and computer setups. The right six grids depict more abstract or symbolic concepts, including lightbulbs, plants with sun motifs, trophies, vehicles, birds, and graduation caps. Each grid contains nine icons arranged in a 3×3 layout, where the top row shows more semantically elaborate versions of the concept, the bottom row shows simpler forms, and icons become visually more abstract from left to right.}
\end{figure*}

Qualitative feedback from our interviews helps explain this preference for \sysname\ and the high satisfaction scores. 
Participants overwhelmingly cited a lack of predictability and control with the baseline, which manifested in both semantic and style properties.
Regarding semantic properties, participants found the baseline's outputs for abstract concepts confusing and unreliable. 
P11 (Female, 23) explained: ``\textit{When I typed in `hope', the results were all over the place. Sometimes it was a candle, then a bird, then a hand...it even gave me a weird combination of all three. I was just left confused about what it was even trying to make.}'' 
This was mirrored in style control. 
Many participants (26/32) described the baseline's behavior as erratic, particularly its failure to maintain consistency in style. 
P29 (Male, 22) expressed his frustration: ``\textit{The (baseline) just did its own thing. I asked it to simplify an outline icon, and it came back as a filled one without me asking. You never knew what it was going to do.}'' As shown in Figure~\ref{fig:user-study}, the baseline's outputs often lack this semantic and stylistic coherence.
This lack of control forced participants into inefficient, frustrating workflows, which explains the low ratings for the outcome satisfaction for the baseline. 
Participants described two strategies when using the baseline: either generating the full grid in a single prompt (Option 2 in Figure~\ref{fig:baseline}), which produced highly redundant and inconsistent results (see P11 in Figure~\ref{fig:user-study}), or adopting a painstaking step-by-step approach to create the nine variations repeatedly (Option 1 in Figure~\ref{fig:baseline}). This manual method imposed a significant cognitive and operational burden. As P30 (Male, 23) put it, ``\textit{Honestly, I was exhausted by the end of that task. It was just this constant loop of thinking and editing.}''
In contrast, participants found \sysname's automated process far more effective. P26 (Female, 22) highlighted the difference: ``\textit{With \sysname, I got to a good place with so much less effort. It just worked.}'' By design, \sysname\ produced icon grids with the clear progressive gradients and stylistic consistency that participants expected.

A qualitative review of the icon grids generated by participants further underscores these differences. 
As shown in the representative examples in Figure~\ref{fig:user-study}, while both systems enabled users to generate icons of varying complexity, \sysname\ demonstrated substantially better consistency and clearer progressive gradients across both semantic and visual dimensions. 
In contrast, the baseline condition often exhibited unexpected style shifts during the simplification process and frequently produced redundant results with poor distinguishability (see P12 and P29 in Figures~\ref{fig:user-study}). 
To further illustrate the capabilities of \sysname\ beyond the study tasks, Figure~\ref{fig:result} presents a gallery of results for 12 additional concepts, including six concrete and six abstract ones.
This visual evidence from both the user study and the gallery confirms that \sysname\ consistently provides the stylistic control and semantic progression that was lacking in the baseline, thereby enabling a more powerful and flexible creative exploration for icon grid design.

\section{Discussion}

In this section, we reflect on the advantages and challenges of \sysname, discuss its limitations, and highlight potential areas for improvement and future research.

\subsection{Scaffolded Progressive Icon Design}

Our research into the \sysname\ system offers a human-AI co-creative approach for icon design, moving beyond the sole reliance on prompt-based generation.
The findings suggest that the potential of generative AI in creative workflows is unlocked not merely by improving the pixel-level fidelity of outputs, but by structurally re-engineering the interaction to mirror the cognitive processes of human creators.
We reflect on the system design of \sysname, focusing on the implications of coupling semantic scaffolds with progressive simplification.

\paragraph{Semantic Scaffolding as a Catalyst for Creative Ideation} 
A primary challenge in visual ideation is overcoming the ``blank canvas'' syndrome, particularly when articulating abstract concepts lacking immediate physical referents.
\sysname\ addresses this by replacing unstructured prompt interaction with semantic scaffolding, breaking down the problem space into manageable pieces.
Previous research (e.g.,~\cite{Wang_Jiang_Cheng_Li_Zhao_2024}) suggests that treating a prompt as a linear sequence of tokens is insufficient for complex, hierarchical tasks. 
Similarly, our scaffold guides the model's attention by structuring the conceptual space before resolving the representational space.
This approach aligns with recent advancements in dimensional scaffolding for design, where structured guidance has been identified as essential for controlling complex, generative outputs~\cite{10.1145/3706598.3714211}. 
By providing this thematic and relational map, \sysname\ ensures structural consistency and functions as a creative partner~\cite{10.1145/3706598.3713500}, augmenting the divergent phase of the creative process rather than merely rendering probabilistic guesses.

\paragraph{Progressive Simplification for Structured Refinement} 
Once a concept is chosen, the creative task shifts from divergence to convergence. \sysname\ supports this phase by computationalizing the manual subtractive process of abstraction, treating it as a constraint-guided synthesis where semantic intent must align with geometric constraints. 
To simulate simplification, \sysname\ combines gestalt generation via SDS and semantic decomposition via SAM, using prioritized pixel merging to consolidate distinct shapes with semantic structures into geometric primitives essential for iconic abstraction~\cite{11094816}. 
This progressive refinement pipeline allows designers to intervene at intermediate abstraction levels, systematically evaluating the trade-off between visual complexity and communicative clarity. 
Technically, by imposing geometric primitives and limiting color palettes, the system steers the latent generation away from the ``Reality'' vertex toward the ``Picture Plane''~\cite{mccloud1993understanding}. 
This approach parallels established methods in engineering drawing vectorization~\cite{1023802}, where progressive simplification is employed to preserve shape characteristics and orthogonality while reducing detail. 
Crucially, \sysname\ extends this principle to the semantic domain, ensuring that the meaning of the icon is preserved even as its information density is reduced.

\paragraph{Navigating Generative Latent Space via Structured Design Exploration}

Current multimodal interfaces often force users to map complex creative intentions onto high-dimensional semantic descriptions, creating a profound ``Gulf of Execution''~\cite{10.5555/576915}. Our evaluation reveals that relying solely on linguistic approximation imposes extraneous cognitive load, trapping users in stochastic, back-and-forth trial-and-error loops to align the model's latent space with their intent.
In contrast, \sysname\ demonstrates that projecting high-dimensional latent spaces onto low-dimensional, deterministic controls (e.g., icon grids) significantly reduces this cognitive tax. 
By enabling users to navigate the design space deterministically, the system fosters creative exploration rather than mere output generation. 
This structured approach moves beyond the disjointed outputs of typical batch generation, directly supporting core design tasks, such as adapting visual abstraction for diverse contexts (e.g., varying between high-abstraction favicons and low-abstraction billboard illustrations). 
Ultimately, effective creative support tools must bridge the gap between generative potential and user control, transforming raw computing power into embodied creative intelligence.

\subsection{Operationalizing Visual Abstraction}

Our evaluation suggests that \sysname\ excels because its technical pipeline mirrors the human cognitive strategy of visual abstraction, situating the system not merely as a tool but as a preliminary computational model. 
In \sysname, we operationalize abstraction not as mere simplification, but as a continuous parameter of semantic density.
We situate our approach within McCloud’s ``The Picture Plane''~\cite{mccloud1993understanding}, traversing the trajectory from ``Reality'' to ``Language''. 
While current generative models excel at the ``Reality'' vertex, they often conflate simplification with degradation (e.g., blurring) or stylization (e.g., filtering). \sysname\ bridges this gap by mapping the model’s high-dimensional latent space to a controllable axis of semantic richness.

This design addresses the inverse relationship between representational and conceptual fidelity in an image, as a photorealistic apple depicts a specific instance (e.g., ``Granny Smith''), whereas a simplified icon communicates the universal concept of ``Apple'' or ``Health''.
\sysname\ enables a deterministic traversal of this concept manifold~\cite{Muttenthaler2025-ng}, allowing designers to move from the specific to the generic while maintaining visual integrity. 
Unlike simple detail minimization, which can yield arbitrary shapes, our system enforces a structural condensation that preserves semantic legibility. 
This effectively externalizes the abstraction process, demonstrating that visual abstraction can be represented not just as a mental cognitive strategy, but as a controllable and reproducible dimension within the computational design space.

\subsection{Harmonizing Human-Machine Co-Creation}

The development of \sysname\ highlights a fundamental friction between human conceptualization and machine computation. 
Human abstraction is a top-down, intentional process grounded in embodied experience, cultural context, and aesthetic sensitivity~\cite{norman2013design, 10.1590/S0102-44502001000100008}. Designers prioritize visual features based on communicative clarity, symbolism, and functional constraints. 
In contrast, machine abstraction in current generative architectures operates through bottom-up statistical optimization, merging pixels based on feature correlations without genuine functional or semantic comprehension~\cite{Ilievski2025-ys}.
This divergence is particularly evident in failure modes, as human error in abstraction can lead to serendipitous creative breakthroughs~\cite{hardy1998creativity}. Machine errors, on the other hand, typically manifest as incoherent statistical artifacts, exposing a lack of genuine intentionality or world knowledge~\cite{Ilievski2025-ys, Muttenthaler2025-ng}.

\sysname\ attempts to bridge this gap by leveraging semantic segmentation to impose a structural prior that mimics human object recognition. 
By enabling users to manipulate semantic scaffolds and simplification levels, the system provides a mechanism to correct ``statistical abstraction'' into ``semantic abstraction'', ensuring outputs align with communicative intent rather than just probability. 
The system computationally replicates the human cognitive strategy of structural condensation by generating a coherent gestalt (via SDS), decomposing it into semantic parts (via SAM), and reconstructing it based on priority.
By aligning its technical pipeline with human cognitive strategies, \sysname\ functions as a scaffold for generation guidance and an amplifier for creative intent.

Consequently, we conceptualize \sysname\ not as an autonomously creative agent, but as a simulator of reductive abstraction that executes a specific strategy with scalability and consistency~\cite{hsu2018categorization}. 
The system functions as a semantic scaffold that organizes the representational choices, augmenting rather than replacing human intentionality.
This integration establishes a collaborative division of labor where machine capabilities structure the creation space by expanding possibilities and handling execution, while the user acts as a creative director. 
This relation leverages the computational capacity for generative breadth alongside the human capacity for aesthetic judgment and contextual decision-making.

\subsection{Limitations and Future Work}

\subsubsection{Limitations}

While \sysname\ demonstrates the promise of AI-assisted icon design, we reflect on limitations that frame the scope of our current contribution.
First, we acknowledge that the insights derived from our formative study are bounded by a relatively small participant sample. While the extracted themes regarding semantic and style control were dominant within this group, a broader inquiry is necessary to uncover latent pain points in other icon design contexts. Future work should extend these insights to a wider demographic of designers to ensure higher ecological validity.

Second, the system's performance is heavily dependent on the initial deconstruction step. While the SAM is robust, segmentation inaccuracies can occur, creating flawed semantic hierarchies that propagate errors downstream. 
To mitigate this, future research could explore interactive ``human-in-the-loop'' mechanisms, allowing designers to correct segmentation boundaries explicitly. Furthermore, our current model of semantic priority relies on heuristics. Future iterations could align more closely with human cognitive models by learning priorities from large-scale design corpora or behavioral signals, such as user gaze patterns.

Finally, in terms of scope, our evaluation compares \sysname\ primarily against a general-purpose workflow (GPT-4o via standard multi-modal chat).
While this baseline demonstrates the capabilities of raw multi-modal generation, it lacks the domain-specific constraints necessary for icon design, particularly in maintaining visual consistency and managing precise abstraction levels.
It also does not capture the complexity of professional workflows involving vector editors and asset libraries.
Therefore, our findings should be interpreted as an assessment of interaction designs atop generative models, rather than a claim that \sysname\ outperforms comprehensive, professional-grade design pipelines.

\subsubsection{Future Work}

While this work demonstrates the potential of generative workflows for coherent icon design, several avenues remain for extending the capabilities of \sysname.
Future research should address the architectural challenge of disentangling semantics from visual style in generative models. While designers mentally separate figurative representation from stylistic attributes, current models often entangle these dimensions, limiting predictable manipulation. We envision advancing model-level representations (e.g., compositional semantic graphs or disentangled latent spaces) to embed abstraction as a controllable internal dimension. This would enable the precise generation of cohesive visual systems, such as brand identities and adaptive pictograms, by allowing designers to alter semantic abstraction without disrupting stylistic consistency.

We also aim to extend abstraction-driven interactions beyond early-stage ideation into full-scale professional workflows. While our current study isolates conceptual exploration, future systems can support the entire design pipeline by integrating abstraction with vector-level editing, multi-resolution adaptation, and style harmonization. 
Bridging the gap between generative exploration and established production environments (e.g., Adobe Illustrator or Figma) would allow for the fluid adaptation of single semantic concepts into multiple variants, ensuring consistency across diverse resolutions and use contexts.

Beyond icon design, the principle of controlled simplification applies to broader visual domains.
In narrative art and comics, this could facilitate character design tools that dynamically slide between realistic and simplified representations to aid reader identification. 
In technical illustration and instructional design, accompanying computer-aided design tools with semantic support could generate diagrams that adapt from abstract schematics to detailed photorealism based on a learner’s expertise. 
Furthermore, this approach holds significant promise for accessible computing, enabling interfaces to dynamically adjust icon complexity based on a user’s visual acuity or cognitive state. 
Ultimately, treating semantic richness as a controllable parameter allows users to manipulate design logic rather than geometry, bridging the gap between functionality and creativity.

\section{Conclusion}

This paper presents \sysname, a human-AI co-creative system designed to address the challenge of creating stylistically consistent icons across spectra of semantic and visual complexity. Our approach organizes the design process along two axes (i.e., semantic richness and visual complexity) and operationalizes this through a novel computational workflow. From a single concept, the system constructs a semantic scaffold to broaden ideation, uses chained, image-conditioned generation to ensure stylistic cohesion, and automatically distills each resulting exemplar into a progressive sequence of representations. These are presented in a navigable grid that makes the creative space explorable.
A within-subjects study with 32 participants demonstrated that, compared to a baseline workflow, \sysname\ enabled the creation of more creative icon grids, significantly reduced cognitive workload, and supported a more coherent exploration of design trade-offs. This work contributes a computational method for progressive icon generation and, more broadly, serves as a proof-of-concept for co-creative systems that structure and scaffold the creative process, bridging human intent with machine capability.

\begin{acks}
We thank the anonymous reviewers for their detailed comments and constructive suggestions that strengthened this work, and all the study participants for their time and valuable feedback.
This research was partially supported by the ICFCRT (W2441020), 
NSFC (U21B2023, 62472288), 
Guangdong Natural Science Foundation (2023B1515120026), 
Shenzhen Science and Technology Program (KQTD20210811090044003, RCJC20200714114435012), 
Israel Science Foundation (3441/21, 1473/24, 2203/24), 
Len Blavatnik and the Blavatnik family foundation, and 
Scientific Development Funds from Shenzhen University.
\end{acks}

\bibliographystyle{ACM-Reference-Format}
\bibliography{main}

\section*{Appendix}

\appendix

\section{Questionnaire Design}

To test our hypotheses, we collected quantitative data comparing \sysname\ and the baseline via a post-task questionnaire. 
This instrument integrated standard scales, adapted measures from the NASA-TLX and CSI frameworks, and custom 7-point Likert scale items to provide a multi-dimensional assessment of the user experience, as detailed below.

\subsection{Cognitive Load}

Q1. \textbf{Mental Demand}: How mentally demanding was this task?

\vspace{-5pt}
\begin{figure}[H]
  \centering
  \includegraphics[width=0.85\linewidth]{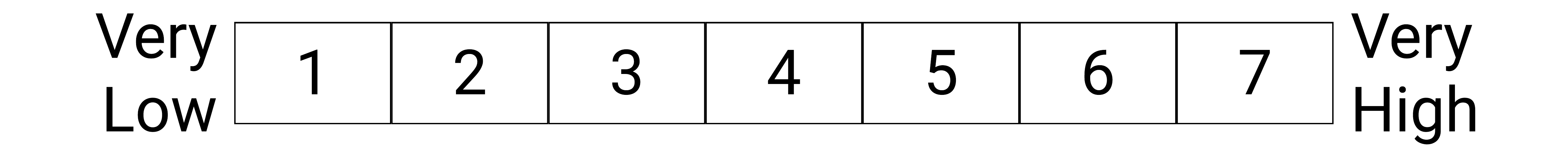}
  \Description{}
  \label{fig:supp-01}
\end{figure}
\vspace{-5pt}

\noindent
Q2. \textbf{Physical Demand}: How physically demanding was this task?

\vspace{-5pt}
\begin{figure}[H]
  \centering
  \includegraphics[width=0.85\linewidth]{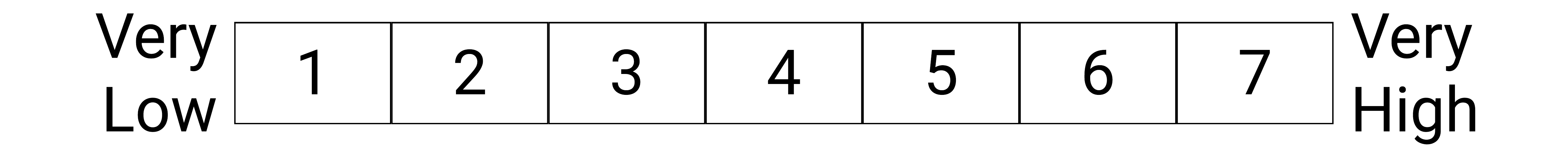}
  \Description{}
  \label{fig:supp-02}
\end{figure}
\vspace{-5pt}

\noindent
Q3. \textbf{Temporal Demand}: How hurried or rushed was the pace of the task?

\vspace{-5pt}
\begin{figure}[H]
  \centering
  \includegraphics[width=0.85\linewidth]{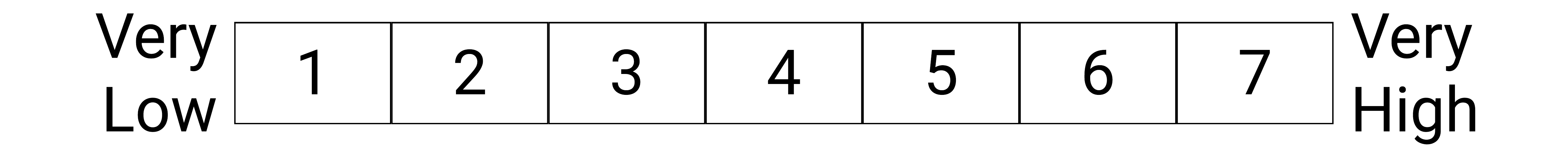}
  \Description{}
  \label{fig:supp-03}
\end{figure}
\vspace{-5pt}

\noindent
Q4. \textbf{Performance}: How successful were you in accomplishing what you were asked to do?

\vspace{-5pt}
\begin{figure}[H]
  \centering
  \includegraphics[width=0.85\linewidth]{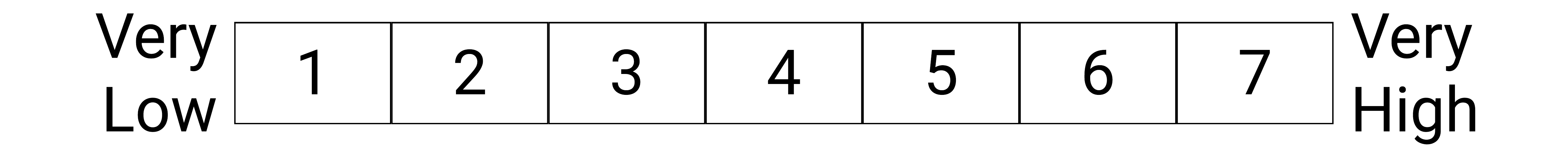}
  \Description{}
  \label{fig:supp-04}
\end{figure}
\vspace{-5pt}

\noindent
Q5. \textbf{Effort}: How hard did you have to work to accomplish your level of performance?

\vspace{-5pt}
\begin{figure}[H]
  \centering
  \includegraphics[width=0.85\linewidth]{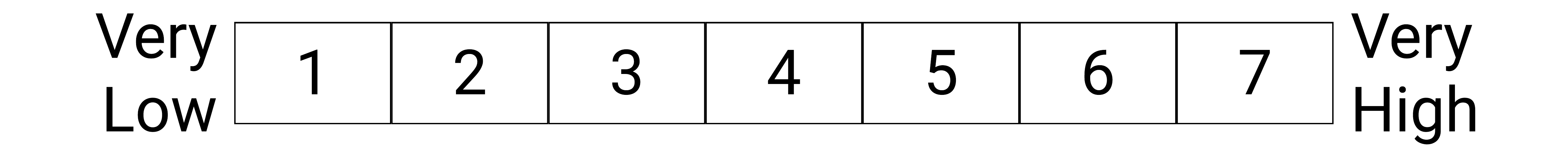}
  \Description{}
  \label{fig:supp-05}
\end{figure}
\vspace{-5pt}

\noindent
Q6. \textbf{Frustration}: How insecure, discouraged, or annoyed did you feel during the task?

\vspace{-5pt}
\begin{figure}[H]
  \centering
  \includegraphics[width=0.85\linewidth]{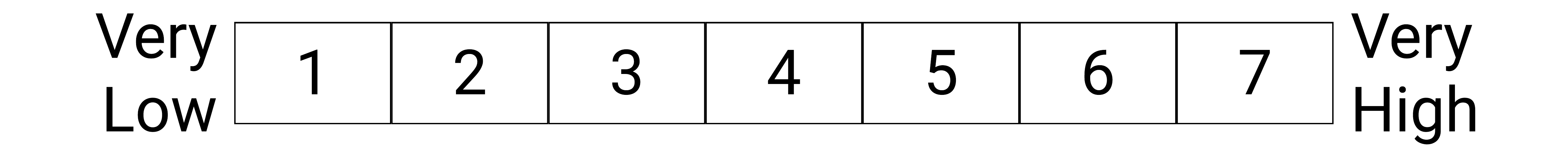}
  \Description{}
  \label{fig:supp-06}
\end{figure}
\vspace{-5pt}

\subsection{Creativity Support}

Q7. \textbf{Exploration}: The system made it easy to explore many different design ideas and options.

\vspace{-5pt}
\begin{figure}[H]
  \centering
  \includegraphics[width=0.85\linewidth]{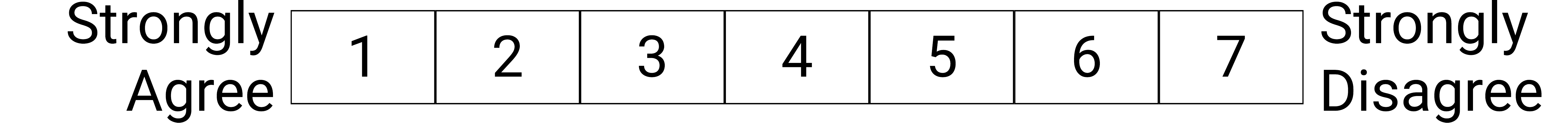}
  \Description{}
  \label{fig:supp-07}
\end{figure}
\vspace{-5pt}

\noindent
Q8. \textbf{Collaboration}: I felt like I was collaborating effectively with the AI.

\vspace{-5pt}
\begin{figure}[H]
  \centering
  \includegraphics[width=0.85\linewidth]{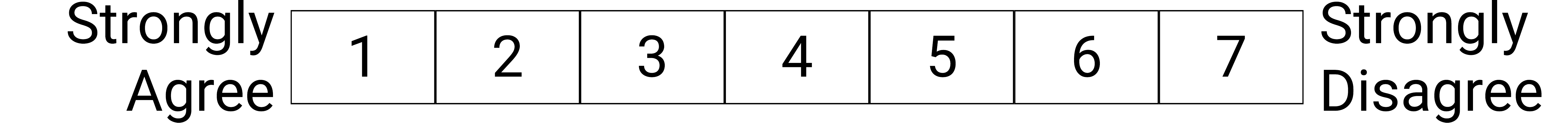}
  \Description{}
  \label{fig:supp-08}
\end{figure}
\vspace{-5pt}

\noindent
Q9. \textbf{Enjoyment}: I found the design process enjoyable and engaging.

\vspace{-5pt}
\begin{figure}[H]
  \centering
  \includegraphics[width=0.85\linewidth]{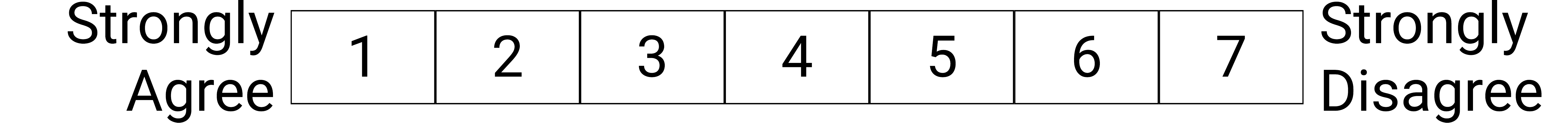}
  \Description{}
  \label{fig:supp-09}
\end{figure}
\vspace{-5pt}

\noindent
Q10. \textbf{Results Worth Effort}: The quality of the icons I created was worth the effort I put in.

\vspace{-5pt}
\begin{figure}[H]
  \centering
  \includegraphics[width=0.85\linewidth]{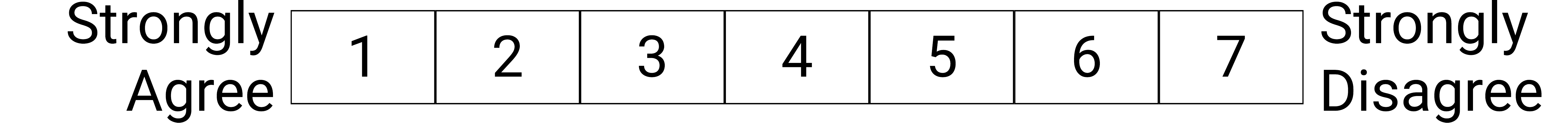}
  \Description{}
  \label{fig:supp-10}
\end{figure}
\vspace{-5pt}

\noindent
Q11. \textbf{Immersion}: The system allowed me to stay focused on my creative task without being distracted by the interface.

\vspace{-5pt}
\begin{figure}[H]
  \centering
  \includegraphics[width=0.85\linewidth]{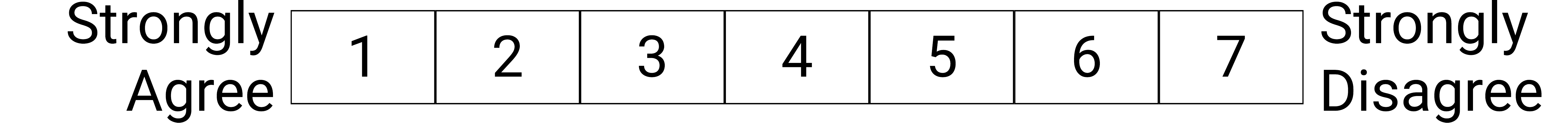}
  \Description{}
  \label{fig:supp-11}
\end{figure}
\vspace{-5pt}

\noindent
Q12. \textbf{Expressiveness}: The system enabled me to be highly creative and expressive.

\vspace{-5pt}
\begin{figure}[H]
  \centering
  \includegraphics[width=0.85\linewidth]{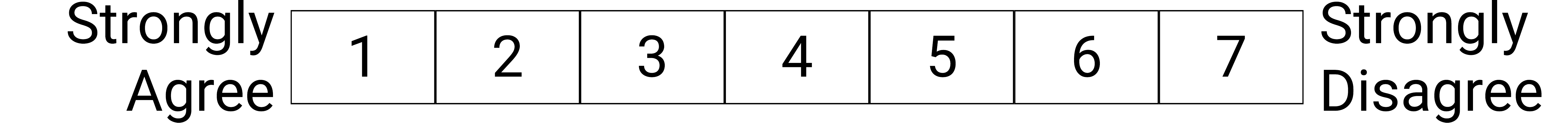}
  \Description{}
  \label{fig:supp-12}
\end{figure}
\vspace{-5pt}

\subsection{Design Satisfaction}

\noindent
Q13. \textbf{Semantic Richness}: The system produced a clear and meaningful range of semantic richness across the icon set.

\vspace{-8pt}
\begin{figure}[H]
  \centering
  \includegraphics[width=0.9\linewidth]{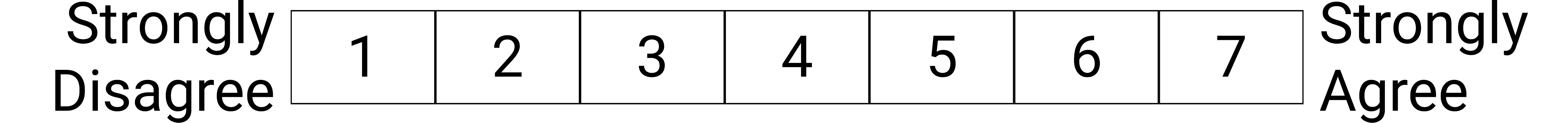}
  \Description{}
  \label{fig:supp-18}
\end{figure}
\vspace{-5pt}

\noindent
Q14. \textbf{Concept Faithfulness}: Please rate your satisfaction with how well the final icons represented your intended concept.

\vspace{-8pt}
\begin{figure}[H]
  \centering
  \includegraphics[width=0.88\linewidth]{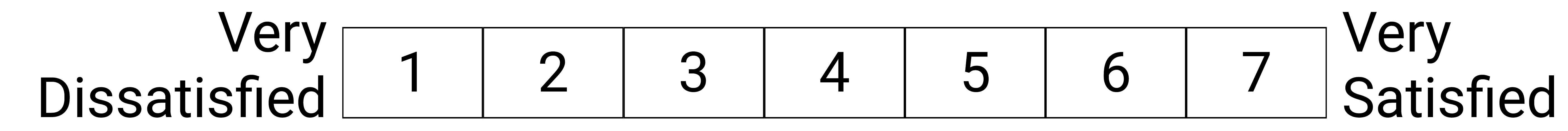}
  \Description{}
  \label{fig:supp-16}
\end{figure}
\vspace{-5pt}

\noindent
Q15. \textbf{Visual Complexity}: The system produced a clear and predictable range of visual complexity across the icon set.

\vspace{-8pt}
\begin{figure}[H]
  \centering
  \includegraphics[width=0.9\linewidth]{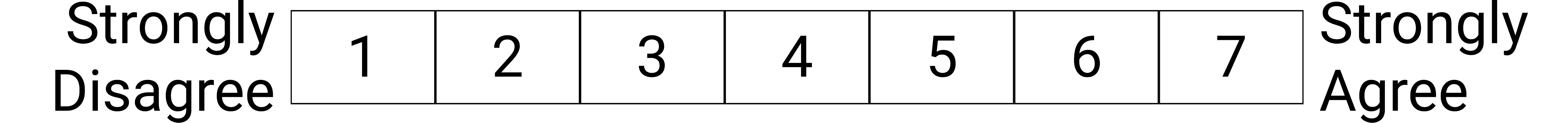}
  \Description{}
  \label{fig:supp-19}
\end{figure}
\vspace{-5pt}

\noindent
Q16. \textbf{Style Consistency}: Please rate your satisfaction with the stylistic consistency of the icons in the final set.

\vspace{-8pt}
\begin{figure}[H]
  \centering
  \includegraphics[width=0.88\linewidth]{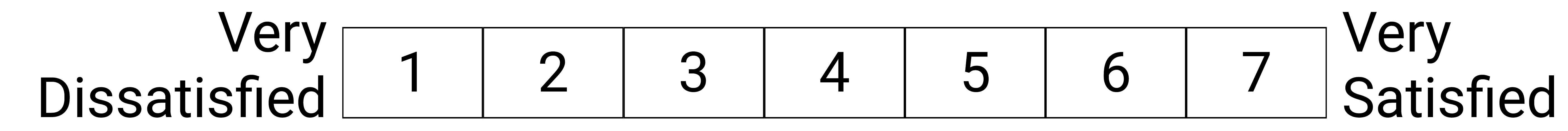}
  \Description{}
  \label{fig:supp-15}
\end{figure}
\vspace{-5pt}

\noindent
Q17. \textbf{Outcome Creativity}: Please rate your satisfaction with the creativity of the icons you produced.

\vspace{-8pt}
\begin{figure}[H]
  \centering
  \includegraphics[width=0.88\linewidth]{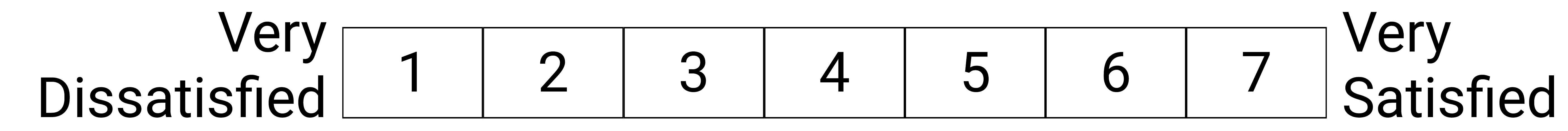}
  \Description{}
  \label{fig:supp-13}
\end{figure}
\vspace{-5pt}

\noindent
Q18. \textbf{Outcome Diversity}: Please rate your satisfaction with the diversity of the icons in the final set.

\vspace{-8pt}
\begin{figure}[H]
  \centering
  \includegraphics[width=0.88\linewidth]{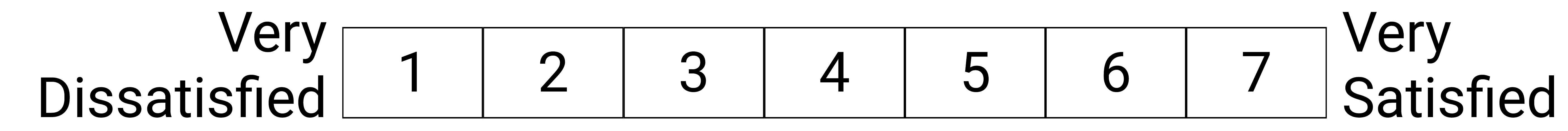}
  \Description{}
  \label{fig:supp-14}
\end{figure}
\vspace{-5pt}

\noindent
Q19. \textbf{Overall Satisfaction}: Please rate your overall satisfaction with the final icon set you created.

\vspace{-8pt}
\begin{figure}[H]
  \centering
  \includegraphics[width=0.88\linewidth]{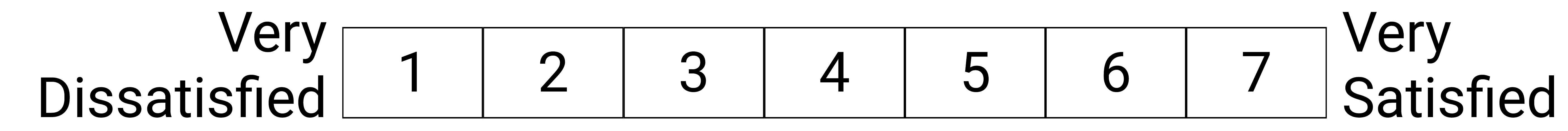}
  \Description{}
  \label{fig:supp-17}
\end{figure}
\vspace{-5pt}

\section{Open-ended Interview Questions for Formative Study}


\subsection*{Part A: Current Icon Design \& Sourcing Workflow} 

\hspace*{\parindent}\textit{Q1. Need for Icons}.  
``In your typical work, what are the most common scenarios or contexts that require you to find or create an icon?''  
\textit{Probe:} ``Are they for user interfaces, presentations, marketing materials, or something else?''

\textit{Q2. The Sourcing Process}.  
``Could you walk me through your process when you need an icon for a specific concept? For example, let's say you need an icon for `Secure Payment'.''  
\textit{Probe:} ``Do you start by designing it yourself, searching a library, or using a generative tool?''

\textit{Q3. Tooling \& Experience}.  
``You mentioned that you used to [design/search/generate] with [tool name]. Could you tell me more about that?''

\begin{itemize}[leftmargin=*, noitemsep]
    \item \textit{If they search libraries:} ``Which libraries or websites do you rely on (e.g., Flaticon, The Noun Project, Material Icons)? What do you like or dislike about the experience of finding the right icon there?''
    \item \textit{If they use AI generators:} ``Which tools have you tried (e.g., Midjourney, Icon-specific tools)? How would you describe the results? How much control do you feel you have?''
    \item \textit{If they design from scratch:} ``What tools do you use (e.g., Figma, Illustrator)? What are the first steps you take when translating a concept into a visual?''
\end{itemize}


\subsection*{Part B: Exploring Challenges \& Unmet Needs}

\hspace*{\parindent}\textit{Q4. Semantic Accuracy vs. Visual Appeal}.  
``When choosing an icon, how do you balance the need for it to be semantically accurate—meaning it clearly communicates the idea—with the need for it to be visually appealing and stylistically consistent?''  
\textit{Probe:} ``Can you recall a time when you had to make a trade-off between these two?''


\textit{Q5. The ``Not-Quite-Right'' Problem}.  
``Have you ever struggled to find or create an icon that perfectly captures the nuance of your intended meaning? Can you share an example where the results you found were close, but not quite right?''  
\textit{Probe:} ``What was missing or wrong about them? How did you end up solving that problem?''


\textit{Q6. The Adaptation Challenge (Level of Detail)}.  
``Icons often need to work across a variety of contexts, from a tiny phone screen to a large monitor or even AR/VR glasses. How do you currently manage this need for adaptation?''  
\textit{Probe:} ``What are the biggest frustrations or time-sinks in this process? Do you feel you lose important details, or does the core meaning become unclear at smaller sizes?''


\subsection*{Part C: Concept Validation \& Feature Ideation}

\hspace*{\parindent}\textit{Introduction Script:}  
``Based on the challenges you've described, we're exploring a new system concept. I'd love to get your initial thoughts on it.''


\textit{Q7. The Core Concept: Progressive Sets}.  
``Imagine a tool where you input a single concept, like `Christmas Tree'. Instead of one icon, it generates a complete, structured set of icons for you. This set would automatically include variations ranging from a very simple, abstract outline to a highly detailed, illustrative version.  
How might a system like this fit into your workflow? How do you see it addressing the challenges we discussed?''


\textit{Q8. Semantic Exploration}. 
``What if, after you input a concept, the system also suggested related concepts to explore—for example, for `Christmas Tree,' it might suggest `ornament,' `pine tree,' `holiday,' or `celebration.'  
Would this be a useful feature for brainstorming or refining your idea? What kinds of related information would be most helpful (e.g., synonyms, broader categories, specific examples)?''


\textit{Q9. Controlling Complexity}.
\begin{itemize}[leftmargin=*, noitemsep]
    \item \textit{Semantic Detail:}  
    ``When choosing between a `tree' vs. a `decorated Christmas tree', what would help you make that decision? How would you want the system to present these semantic levels?''
    \item \textit{Visual Detail:}  
    ``When choosing between different visual styles (e.g., line thickness, color, decorative elements), what specific attributes are most important for you to be able to see and control?''
\end{itemize}


\textit{Q10. The Ideal Tool}.  
``If you had a magic wand and could design the perfect icon creation tool to solve these problems, what would it do for you?''


\subsection*{Part D: Wrap-up}

\noindent ``Those are all my questions. Is there anything else you'd like to add about your experience with icons, or any questions you have for me?''

\section{Open-ended Interview Questions For User Study}

\paragraph{Q1. Suggestions for System Improvement} 
``What aspects of the system or generated results would you like to see improved, and how? How do you assess the system's performance for control with the automation of progressive simplification?''

\paragraph{Q2. Importance of Semantic Content vs. Visual Style} 
``In icon design, how do you distinguish between style and content? Which do you consider more important: the subject matter (e.g., a camera, a person) or the visual style (e.g., simple, cartoonish), and why?''

\paragraph{Q3. Preferences for Stylistic Variations} 
``What are your suggestions regarding the generation of specific styles (e.g., outline, filled, color)? Please describe your preferred styles, the contexts in which you use them, and any additional styles you would like to see.''

\begin{figure*}[t]
  \centering
  \includegraphics[width=\linewidth]{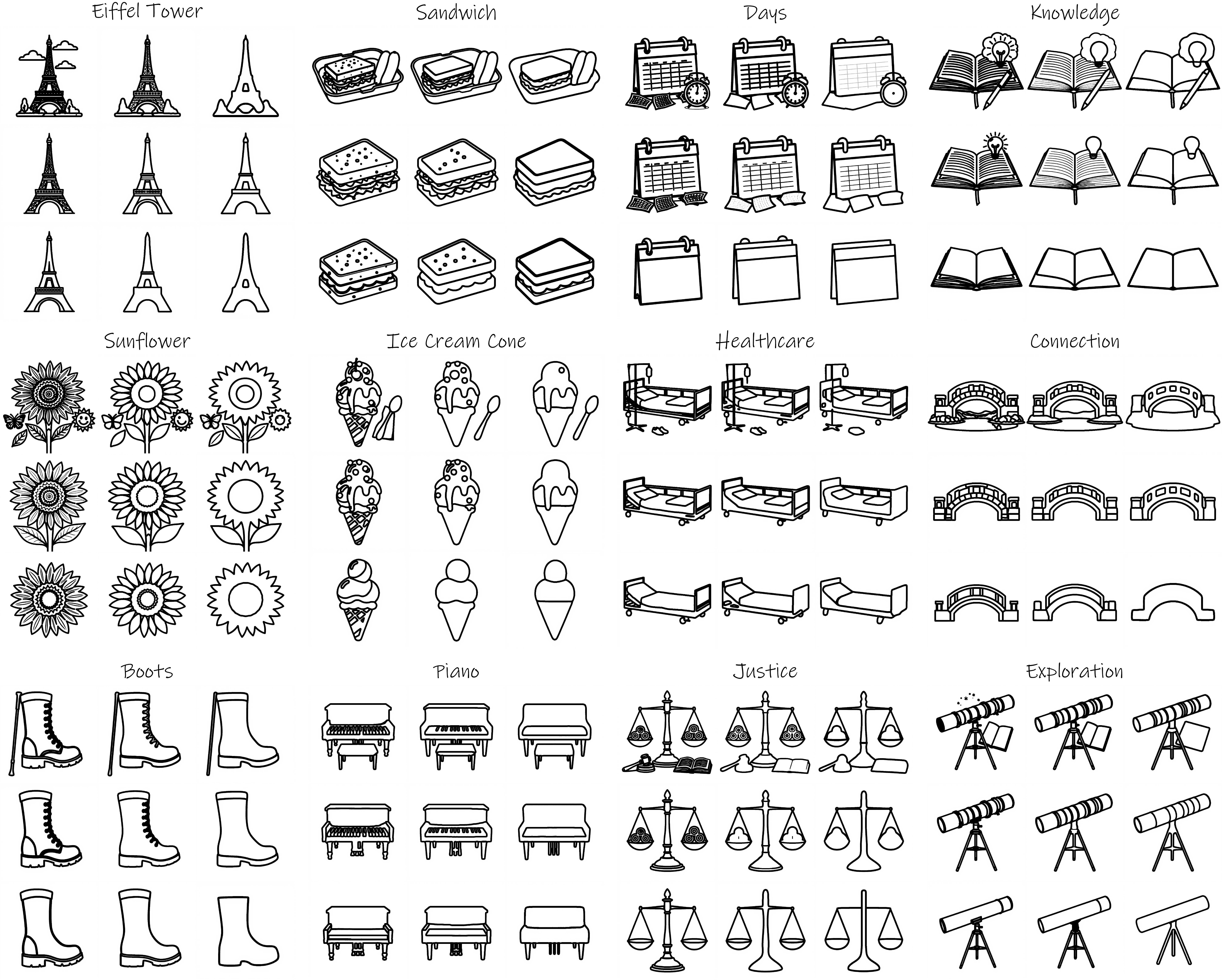} 
  \caption{A gallery of additional icon sets generated by \sysname. The examples demonstrate the system's ability to produce diverse and coherent outcomes for a range of concrete (left) and abstract (right) concepts, which were selected from prior research and common use.}
  \Description{More examples.}
  \label{fig:supp}
\end{figure*}

\paragraph{Q4. A Computational Model for Visual Abstraction}
``In your observation, do you define abstraction primarily as a geometric reduction (e.g., minimizing the number of strokes or edges) or as a semantic preservation task (e.g., maintaining recognition of specific object parts)? How does your model decide which visual details are `essential' and which are `noise'?''

\paragraph{Q5. On the Nature of Human vs. Machine Abstraction}
``In your observations, where does the machine's abstraction logic most significantly diverge from human intuition? Are there specific `errors' the model makes that a human designer would never make?''

\section{STANDARDIZED SEED PROMPT} 

\begin{Verbatim}[breaklines=true]
You are an icon design assistant. Your task is to analyze the user’s input concept and generate an icon. Please follow the workflow below:

[Workflow]
1. Interpret the user’s request and identify the concept they want to design an icon for.
2. Ask for or confirm the two control dimensions based on the user’s input:
   - Semantic Richness
   - Visual Complexity
3. Generate an icon description that strictly adheres to these two dimensions.

[Control Dimension Definitions]
1. Semantic Richness: Determines how many semantic elements are included in the icon.
   - High: Incorporates multiple semantically related elements, symbols, or 
     features.
   - Low: Includes only the most essential and core concept elements.

2. Visual Complexity: Determines the level of visual detail in the rendering.
   - High: Realistic, detailed, natural/organic shapes.
   - Low: Abstract, geometric, minimalistic, symbolic.

[Your Task]
For any concept the user provides, you must produce an icon description that matches the specified Semantic Richness and Visual Complexity. You should guide the user through the design process, following the workflow above.
\end{Verbatim}

To support reproducibility and encourage further research in generative icon design, we have made our materials open-source.
These resources are available at https://github.com/Young-Allen/Iconix.

\section{More \sysname: Additional Cases}

To further demonstrate the versatility and creative capacity of \sysname, Figure~\ref{fig:supp} presents a gallery of 12 additional icon sets generated by the system. 
These examples highlight the system's ability to produce diverse and coherent outcomes for a range of both concrete and abstract concepts.

\end{document}